\documentclass{article}
\usepackage{chemformula}[2014/04/08]
\usepackage{authblk}
\usepackage{tabularx}
\usepackage[english]{babel}
\usepackage{siunitx}
\DeclareSIUnit\angstrom{\text {Å}}
\usepackage[letterpaper,top=2cm,bottom=2cm,left=3cm,right=3cm,marginparwidth=1.75cm]{geometry}

\usepackage{amsmath}
\usepackage{graphicx}
\usepackage[colorlinks=true, allcolors=blue]{hyperref}

\title{\textbf{Chromium-doped uranium dioxide fuels: A review}}
\author[1,2]{Mack Wesley Cleveland}
\author[3]{Andrew Nelson}
\author[1,4,5*]{Ericmoore Jossou}
\affil[1]{Materials in Extreme Environment Laboratory, Massachusetts Institute of Technology (MIT), 77 Massachusetts Ave., Cambridge, 02139, MA,USA}
\affil[2]{Department of Materials Science and Engineering, MIT Cambridge, MA, USA}
\affil[3]{Oak Ridge National Laboratory, Oak Ridge, TN, United States}
\affil[4]{Department of Nuclear Science and Engineering, MIT Cambridge, MA, USA}
\affil[5]{Department of Electrical Engineering and Computer Science, MIT Cambridge, MA, USA}
\affil[*]{Corresponding author: \href{ejossou@mit.edu}{ejossou@mit.edu}}

\begin{document}
\setchemformula{kroeger-vink}
\maketitle

\begin{abstract}
$\ch{UO_2}$ doped with parts per million $\ch{Cr_2O_3}$ powder is considered a potential near term accident tolerant fuel candidate. Here, the results of decades of industry and academic research into Cr-doped $\ch{UO_2}$ are analyzed and their shortcomings are critiqued. Focusing on the incorporation mechanisms of Cr into the fuel matrix, we explore a mechanistic understanding of the characteristic properties of Cr-doped $\ch{UO_2}$, notably, enhanced fission gas retention attributed to enlarged grain sizes following sintering, along with marginal improvements in the thermophysical properties. The findings of recent X-ray Adsorption Near Edge Spectroscopy studies were compared and put into conversation with historic data regarding the incorporation of Cr in $\ch{UO_2}$. On the basis of defect mechanisms, the case is made for the substitutional incorporation of Cr governing the lattice solubility but not the enhanced U diffusivity. Instead, Cr/$\ch{Cr_2O_3}$ redox chemistry in a well-defined oxygen potential explains the differences in the U diffusivity and O/M ratio. The primary mechanism of doping enhanced grain growth is found to be liquid assisted sintering due to a $\ch{CrO_{(l)}}$ eutectic phase at the grain boundaries.  The role of inhomogeneities in Cr concentration in $\ch{UO_2}$ at various length scales across the materials microstructure is highlighted and connected to promising experimental and modeling work to fill in the gaps in the current understanding of Cr-doped $\ch{UO_2}$. The review considers both the open scientific questions and engineering applications to illustrate the deep connections between the practice and theory in the design of accident tolerant nuclear fuels. The review ends with an outline of future works that combine meticulous irradiation studies and high resolution experiments with next generation modeling and simulations techniques empowered by machine learning advances to accelerate the fabrication and adoption of Cr-doped $\ch{UO_2}$ light water reactors. 
\end{abstract}

\section{Introduction}
Uranium dioxide ($\text{UO}_2$) is the primary fuel used in modern light water reactors (LWRs). The use of $\text{UO}_2$ in nuclear reactors dates back to the operations of the Chicago Pile-1 reactor during the Manhattan Project \cite{Belle_UO2_1961}. Meanwhile, its use in commercial nuclear power reactors can be traced back to 1957, within three years of the Atomic Energy Act of 1954 enabling civilian power \cite{sadik_evaluation_1959}. It has been the dominant nuclear fuel material since 1961 \cite{hewlett_new_1962}. The suitability of $\text{UO}_2$ as a nuclear fuel stems from its chemical, thermal, and radiation stability \cite{LYONS1972167,mceachern_review_1998}. These properties are weighed favorably against its lower thermal conductivity and U density than alternative metallic and other ceramic fuels \cite{FINK20001} as safe reactor operation supersedes other concerns.  

Due to fuel fragmentation, relocation, and dispersion (FFRD) concerns, fission gas release (FGR) is a performance limiter of $\ch{UO_2}$ fuels in LWR operations. The nuclear fission of the fuel atoms produces radionuclides, including inert gases such as He, Kr, and Xe, that are insoluble in the $\text{UO}_2$ lattice. These fission products eventually migrate to sink sites such as grain boundaries (GBs) \cite{kolstad_fuel_1992}. At the GBs, they can form bubbles, causing the swelling, embrittlement, and reduced thermal conductivity of the fuel \cite{COOPER2021152590}. The FGR from the fuel, along with the burst behavior of the fuel cladding together, limits the lifetime of the fuel before replacement  \cite{rest_fission_2019}. 

Steady state fission gas release is generally governed by the time it takes fission gas to diffuse through the $\text{UO}_2$ lattice to the grain boundaries  \cite{kolstad_fuel_1992, COOPER2021152590}. Increasing the grain size means that fission gas has to travel further to reach grain boundaries, making the rate of steady state FGR proportional to $1/r^2$ where $r$ is the grain size \cite{BOOTH1957}. An economically efficient way of increasing the grain size of the nuclear fuel material would thus slow FGR and increase the safe reactor operation time to enable higher burn-up. 

The fuel pellets are typically manufactured from precursor powders that are sintered to increase grain size and density of the fuel. The common standard base purity specification for sinterable $\ch{UO_2}$ powder is ASTM C753-16a, which specifies that the concentrations of impurity elements can be no greater than 100-300 weight parts per million (wppm) individually and 1500 wppm in total \cite{ASTM_C753-16}. The introduction of the standard expects buyers to supplement the specification with their proprietary modifications, a practice consistent with the common practice of including additives in commercial grade $\ch{UO_2}$ pellets. During sintering, the powder is heated to a temperature above \SI{1700}{\celsius} ($0.64 T_m$)\cite{HARADA1997217} so that diffusion is faster and grain growth can be promoted \cite{BURKE1952220}. There are numerous studies\cite{doi1964significance,https://doi.org/10.1111/j.1151-2916.1963.tb11692.x} of this sintering process, including the remarkable finding that impurities intentionally included in the precursor powder can accelerate grain growth \cite{kashibe1998effect,Winter1965} as shown in Table \ref{tab:sintering_params}. The doping of $\text{UO}_2$ with concentrations between 100 and 10,000 parts per million (ppm) of metal oxides provides a cost effective way to increase the grain size of $\text{UO}_2$ to improve the fission gas (FG) retention of the fuel \cite{etde_20628373}. 

Because larger grained $\text{UO}_2$ leads to lower FGR \cite{TURNBULL197462}, thus improved margins of safety during accident conditions, doped $\text{UO}_2$ is considered in the class of accident tolerant fuels that received renewed attention following the accident at Fukushima Daiichi Power Plant \cite{zinkle_accident_2014}. A large inventory of historical data and chemical similarity of undoped and doped $\text{UO}_2$ fuel also facilitate  a shorter path to regulatory approval and a higher technological readiness level than other accident tolerant fuel forms. Over the past decades, with additions of $\text{Cr}_2\text{O}_3$ and sometimes codoping with $\text{Al}_2\text{O}_3$, the certification process has been shortened through extensive academic and industry research efforts. Cr-doped $\text{UO}_2$ is already used in reactors in Europe \cite{key} and has received approval for deployment in pressurized water reactors in the United States \cite{ADOPT_NRC}. 

Therefore, the present article aims to look at the current understanding of Cr-doped $\text{UO}_2$ to assess what we know and what gaps remain in our knowledge towards deployment in commercial reactors. This review article is organized into five sections that includes: (i) Section 2 gives an overview of the historical use of dopants in uranium dioxide (ii) Section 3 presents the current status on the solubility limits of Cr doping and the thermodynamic limitations during sintering  (iii) Section 4 discusses the emerging properties due to the addition of Cr to $\text{UO}_2$, (iv) Section 5 is focused on the experimental and first principles calculations efforts towards the understanding how Cr incorporation into $\text{UO}_2$ lead to accelerated grain growth and finally, (v) Section 6 puts the state of the arts in perspective and offers some suggestions for future work. Therefore, through a critical survey of the literature, we show what areas we can be confident in the properties of Cr-doped $\text{UO}_2$ and where uncertainties remain. By searching for a mechanistic understanding, the present review offers insights to guide the incremental optimization of nuclear fuel materials, including other candidate dopants. 

\section{Historical perspectives on the use of dopants in uranium dioxide}

The use of additives and secondary phases to enhance the performance of $\ch{UO_2}$ has been prevalent since the early decades of the nuclear industry. The superior irradiation behavior of $\ch{UO_2}$ compared to U metal resulted in its widespread adoption as an early reactor fuel, but its low thermal conductivity was acknowledged as a significant drawback to its use as a high-power reactor fuel \cite{Lewis1957}. This prompted studies where high thermal conductivity, often metallic, secondary phases were added to $\ch{UO_2}$ a volume typically greater than ten percent. Research into multiple classes of $\ch{UO_2}$-based composite fuel forms with this objective continues in the modern era \cite{TERRICABRAS2025156114, MIDDLEBURGH2023154250, Murphy2024}. A secondary area of focus for early nuclear fuel developers was the use of ceramic additives to modify the microstructure and properties of $\ch{UO_2}$. Although unsuccessful, very early work in this area focused on attempts to use secondary cations to modify the electronic structure of $\ch{UO_2}$, with the stated goal of increasing the electronic contribution to thermal conductivity to meaningful levels \cite{Powers1960}. 

Work initiated during the same time focused on the challenges introduced by industrial processing of $\ch{UO_2}$ fuel pellets. A 1960 manuscript cites the challenges pure $\ch{UO_2}$ feedstock encounters in achieving uniform microstructure and properties without resorting to high cold pressing pressures and sintering conditions viewed as problematic for commercial scale processes of the era \cite{ANG1960176}. The work found that CaO and $\ch{TiO_2}$ additives included at 0.25 weight percent were successful in improving the sintering behavior and reducing the dwell temperature needed to achieve acceptable density pellets. This outcome demonstrated for the first time in the open literature the potential for additives to impart favorable effects on the processing and microstructure of $\ch{UO_2}$. 

These initial efforts to evaluate the potential of $\ch{UO_2}$ dopants grew in multiple directions in the 1960s and 1970s. Titanium dioxide remained the most prevalent additive studied in the open literature during this era, with numerous processing-structure-property studies published. The emphasis of research during this period was development of improved large scale processing routes to enable industrial fabrication of $\ch{UO_2}$. The thermochemistry of the U-O system presents a challenge to commercial sintering: reducing atmospheres retard sintering kinetics, but oxidizing atmospheres introduce volatility of the vapor phase $\ch{UO_3}$. Titanium dioxide was identified as an additive to overcome this limitation and was found to be a viable means to enhance sintering kinetics and encourage grain growth \cite{ainscough_rigby_osborn_1974, AMATO1966252}. 

Use of sintering additives in this manner has been employed by the ceramics industry for centuries, with multiple potential outcomes depending upon the microstructure-performance requirements of the material \cite{German01042013}. Secondary phases or secondary cations into the feedstock can induce numerous impacts on the sintering behavior and resultant microstructure as dictated by the thermochemistry of the system \cite{hannay2012treatise}. In the case of $\ch{TiO_2}$ added to $\ch{UO_2}$, the solubility of Ti 3+/4+ in the $\ch{UO_2}$ lattice greatly increases specie diffusivity up until the solubility limit is exceeded. At higher $\ch{TiO_2}$ contents, precipitation of a secondary phase induces the opposite effect, pinning grain boundaries \cite{PAIK201334}.   

Early research explored numerous other additives to $\ch{UO_2}$ with a similar objective. The patent literature of the era demonstrates the migration of this research from the laboratory to commercial interests. American, British, and Japanese intellectual property cite $\ch{TiO_2}$  in addition to other oxides as additives to $\ch{UO_2}$ fuel pellets for commercial processes with the stated goal of improving sinterability and homogenizing the microstructure \cite{googleUS4430276AMethod,googleGB1285190AImprovements, toraji1968method}. Studies then began to examine the role that additives may have on irradiation performance. The first rigorous theoretical treatment of how cation doping of $\ch{UO_2}$ may impact fission gas behavior was published by Matzke in 1966 \cite{matzke_atomic_1966}. Matzke connected the interplay of cation doping in the $\ch{UO_2}$ lattice on sintering kinetics and grain growth, and further extended these effects to hypothesized impacts on the diffusivity of fission gas atoms. These insights facilitated subsequent development of models correlating fission gas release to grain size as outlined in the previous section, and moreover provided the first indications that dopants present in the $\ch{UO_2}$ lattice may increase the rate of fission gas diffusion in the matrix \cite{UNE198793}.

Correlation of how dopants may impact not only the processing of $\ch{UO_2}$ but also the fuel performance prompted the modern era of dopant research. Given the incremental nature of dopant impact on fuel performance and licensure, much of the full body of research remains proprietary. Commercial investigation of dopants’ use with the stated objective of improving fuel performance is evidenced in the patent literature as early as 1988 \cite{googleFR2706066B1Nuclear}. Academic investigations of dopant impacts on $\ch{UO_2}$ processing, properties, and performance are found in the literature throughout recent decades, with multiple examples cited throughout the following sections. 

\section{Cr incorporation into uranium dioxide microstructure}
Commercial scale fabrication methods of Cr-doped $\text{UO}_2$ powders are under development, but more data exists on preparation of Cr-doped $\text{UO}_2$ in the laboratory setting. Cr doped $\text{UO}_2$ refers to materials prepared by mixing $\text{UO}_2$ powder with small amounts of chromia ($\text{Cr}_2\text{O}_3$) or metallic Cr powder. The resulting powder blend is sintered to increase the grain size compared to the undoped $\text{UO}_2$. In preparing the mixed powder, the Cr-containing precursor is often mechanically mixed to ensure homogeneous distribution of the dopants, or else a wet synthesis route is used to obtain an even Cr concentration throughout the sample. The Cr dopants may fully dissolve in the $\text{UO}_2$ bulk or lattice, segregate to areas like the surface or grain boundaries, or precipitate as a secondary phase. This section presents the current understanding of each of these behaviors in doped $\text{UO}_2$ fuel.

\subsection{Lattice solubility}\label{solubility}
\begin{figure}
    \centering
    \includegraphics[width=1.0\linewidth]{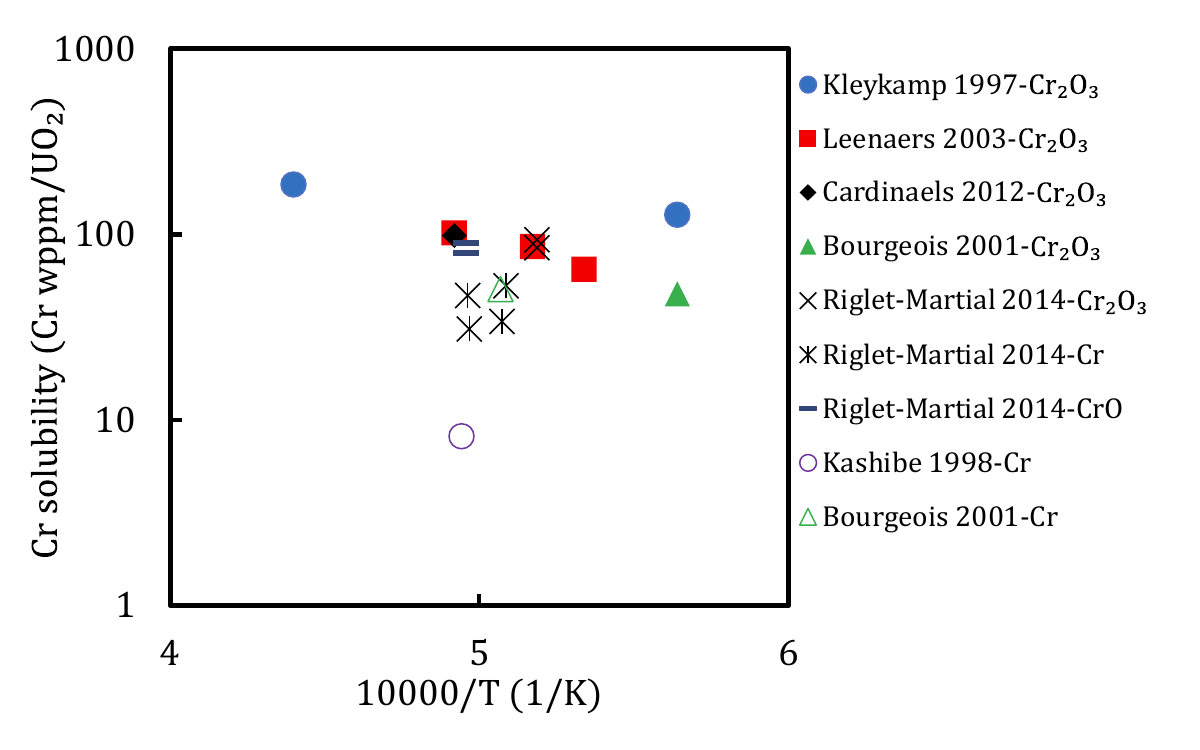}
    \caption{Cr solubility in $\text{UO}_2$ as a function of temperature from literature \cite{KLEYKAMP1997103,leenaers2003solubility,cardinaels2012chromia,BOURGEOIS2001313,RIGLETMARTIAL201463,kashibe1998effect}. Stable phase of Cr-O phase diagram for sintering conditions is annotated in the legend. All but Bourgeois \textit{et al.} used EPMA to characterize their Cr content after sintering, whereas Bourgeois reported the minimum Cr composition at which grain growth slowed due to solute drag. Solubilities are in wppm Cr per $\text{UO}_2$ rather than the expected phase.}
    \label{fig:solubility}
\end{figure}

Bourgeois \textit{et al.} first proposed a solubility limit of 700 ppm at \SI{1700}{\celsius} for  $\text{Cr}_2\text{O}_3$ in $\text{UO}_2$ on the basis of the concentration at which the grain growth was hindered by solute drag \cite{BOURGEOIS2001313}. Leenaers \textit{et al.} were the first to directly measure the solubility of Cr in the bulk using electron probe microanalysis \cite{leenaers2003solubility}. They measured the maximum Cr solubility at \SI{1760}{\celsius} to be 1750 ppm $\text{Cr}_2\text{O}_3$ and also observed that the Cr solubility is proportional to the temperature, as shown in Figure \ref{fig:solubility}. Riglet-Martial \textit{et al.} proposed a thermodynamic model of Cr solubility in $\text{UO}_2$ \cite{RIGLETMARTIAL201463}. The slope of the log solubility versus log oxygen partial pressure is proportional to the stoichiometric coefficient of oxygen needed to oxidize Cr in the thermodynamically stable phase into the $\text{Cr}^{+3}$ oxidation state and fits the existing electron probe microanalysis data. As a result, Riglet-Martial showed that the solubility is proportional not only to temperature but to the oxygen potential within the domain of $\text{Cr}_2\text{O}_3$ stability, which presents the ultimate solubility limit for Cr for a given sintering temperature. Therefore, the difference between Leenaers and Bourgeois is explained by the disparity in the oxygen potential of their sintering environments. The thermodynamic model proposed by Riglet-Martial is in good agreement with the ensuing experimental work and has been effectively used to determine solubility limits \cite{milena2021raman, MURPHY2023}. Notably, the solubility model by Riglet-Martial assumes that Cr is incorporated into the lattice in its +3 charge state. Figure \ref{fig:solubility} shows the measured solubility as a function of temperature taken from the literature. The specimens sintered in the domain of $\ch{Cr_2O_3}$ stability show a direct dependence on temperature, with solubility being proportional to temperature. However, the specimens sintered in the domain of metallic Cr show a larger dependence on O potential following the model of Riglet-Martial \textit{et al}.

Prediction of Cr solubility in $\text{UO}_2$ from simulation has encountered mixed success with some first principles calculations showing positive solution energies for Cr defect complexes in $\text{UO}_{2-x}$, though solution is predicted to be more favorable in $\text{UO}_{2+x}$ where U vacancies are available to be eliminated to provide the charge compensation and sites for Cr substitutions \cite{MIDDLEBURGH2012258,Gascoin2025}. Moreover, density functional theory calculations from Middleburgh \textit{et al.} support the claim that Cr incorporates as a +3 state in the lattice, taking an electron from a neighboring U, oxidizing it to a +5 charge state. Moreover follow up modeling work using empirical interatomic potentials found that Cr preferentially form a $\ch{CrUO_4}$ phase \cite{COOPER2013236} although this has not been confirmed by experiment \cite{MURPHY2023} pointing to the limitations of empirical potentials. Cooper \textit{et al.} also used first principles DFT+U energies and vibrational entropies from empirical potential to predict the equilibrium Cr solubility at an oxygen partial pressure of $10^{-20}$ \SI{}{atm} \cite{COOPER2018403}. Cooper found that at high temperature and reducing environments, Cr in a reduced oxidation state of $\text{Cr}^{+1}$ is almost completely soluble at interstitial sites due to an increase in defect volume for higher Cr valence that produced a corresponding increase in defect vibrational entropy when applying a volume correction to the vibrational entropy obtained with fixed +3 charge state using $S_{\ch{Cr}^q}=S_{\ch{Cr}^{+3}}+(V_{S_{\ch{Cr}^q}}-V_{S_{\ch{Cr}^{+3}}})\dot (dS/dV)_{\ch{UO_2}}$. This result contradicts the experimental solubility model of Cr, in which case Cr is incorporated substitutionally, and the solubility is proportional to oxygen partial pressure. Cleveland and Jossou recently extended Cooper \textit{et al.}'s approach to the domain of experimental sintering conditions and found that Cr incorporates substitutionally at concentrations in reasonable agreement with the experimental solubility limits \cite{ClevelandChargestate}. Therefore, first principles prediction of the solubility of Cr in $\text{UO}_{2}$ provides support for the experimentally based thermodynamic models while also suggesting several viable possibilities outside the thermodynamic domain currently considered in experiments. 

\begin{figure}
    \centering
    \includegraphics[width=1.0\linewidth]{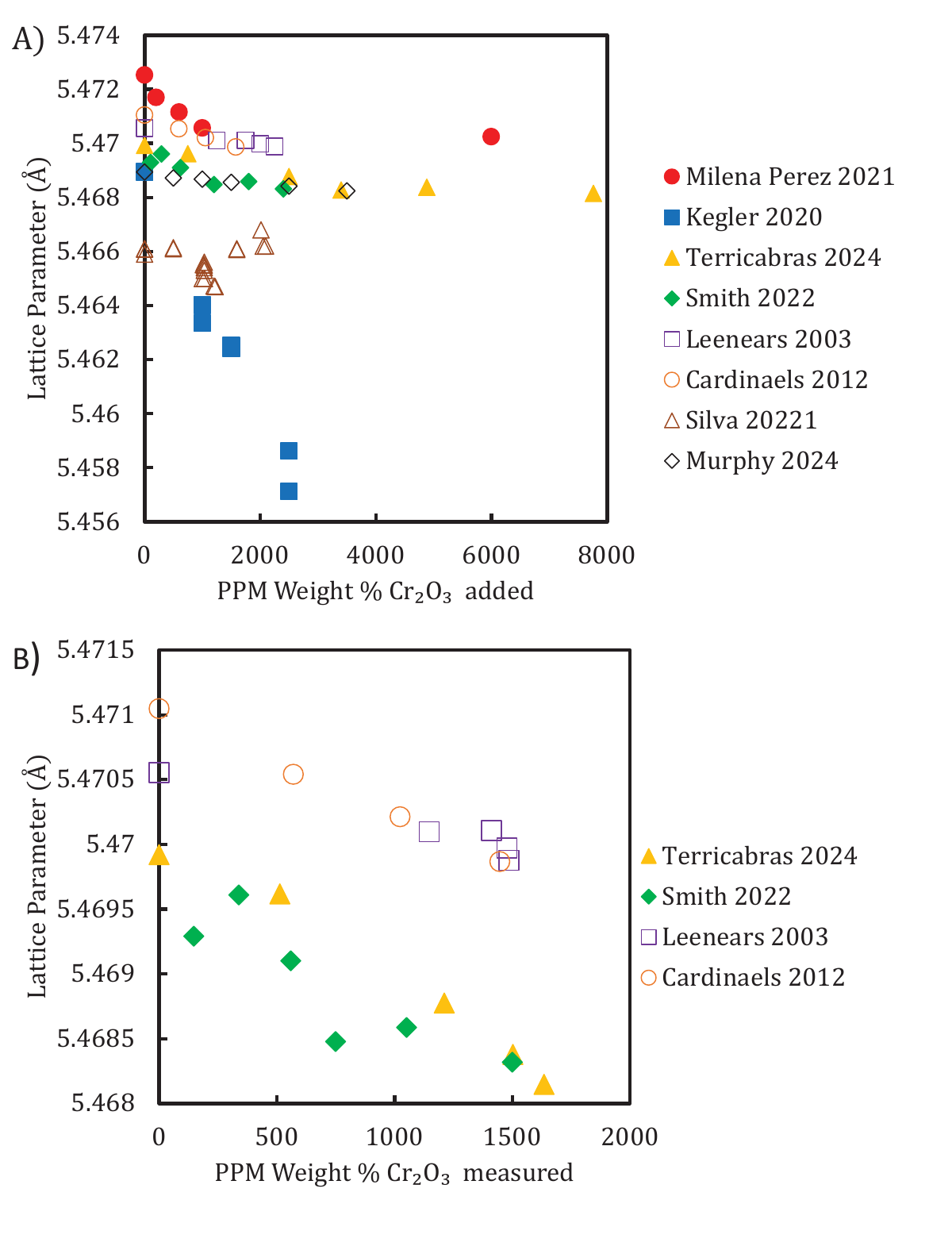}
    \caption{Lattice parameter of Cr-doped $\text{UO}_2$ (a) versus weight percent $\text{Cr}_2\text{O}_3$ added initially, (b) versus measured concentration of $\text{Cr}_2\text{O}_3$ in weight percent equivalent. Values taken from \cite{milena2021raman,kegler2020chromium,TERRICABRAS2024155022,SMITH2022,leenaers2003solubility,CARDINAELS2012289,silva2021evaluation,Murphy2024} }
    \label{fig:Lattice parameter}
\end{figure}

Experiments have shown that the lattice parameter of $\text{UO}_2$ contracts with increasing additions of $\text{Cr}_2\text{O}_3$ dopants, as shown in Figure \ref{fig:Lattice parameter}. Leenaers \textit{et al.} first observed the lattice contraction with X-ray Diffraction (XRD) of Cr doped $\text{UO}_2$ powders, but concluded that the magnitude of the lattice contraction was too small to be accounted for by a substitutional defect, which they hypothesized would more strongly distort the lattice \cite{leenaers2003solubility}. Instead, they proposed an interstitial mechanism that would lead to cation vacancies. A subsequent study by Cardinaels \textit{et al.} validated their own observed lattice contraction with results using an empirical potential for Cr substitutions, but found that the defect clusters slightly overestimated the reduction in lattice parameter \cite{cardinaels2012chromia}. The contraction in the lattice parameter is observed by subsequent studies \cite{SILVA2021153003,kegler2020chromium,SMITH2022} , though with quantitative differences in both the initial lattice parameter and the contraction. Kegler \textit{et al.}\cite{kegler2020chromium} using a wet synthesis method report the largest variation in lattice parameter, whereas Silva \textit{et al.} \cite{silva2021evaluation} report almost no variation in lattice parameter. This points out the limitation of studies that report only the initial concentration of $\text{Cr}_2\text{O}_3$ added rather than the concentration of Cr in the lattice after sintering because of the volatilization of Cr dopants. Terricabas \textit{et al.} \cite{TERRICABRAS2024155022} and Milena-Perez\textit{et al.} both looked at the effect of doping above the lattice solubility limit and found a change in the slope of lattice contraction with no additional lattice contraction after about 2500 ppm $\text{Cr}_2\text{O}_3$ addition to the $\text{UO}_2$ matrix. However, Terricabas \textit{et al.} showed that the lattice contraction is proportional to the Cr concentration in the lattice, as shown in Figure \ref{fig:Lattice parameter} B). This explains the apparent difference between these results and a more recent study by Murphy \textit{et al.} using synchrotron X-ray diffraction that found a lower lattice contraction even above the lattice solubility limit, which is proposed to be due to a secondary Cr phase \cite{Murphy2024}. While there is still variation between studies for the slope of lattice contraction versus actual lattice Cr content \cite{leenaers2003solubility,cardinaels2012chromia,SMITH2022,TERRICABRAS2024155022}, the variability in the slope in Figure \ref{fig:Lattice parameter} B) is reduced compared to the nominal in Figure \ref{fig:Lattice parameter} A). Accordingly, the evidence points to a small contraction in the $\text{UO}_2$ lattice due to the incorporation of Cr ions substituting for U in the $\text{UO}_2$ bulk lattice. 

\begin{figure}
    \centering
    \includegraphics[width=1.0\linewidth]{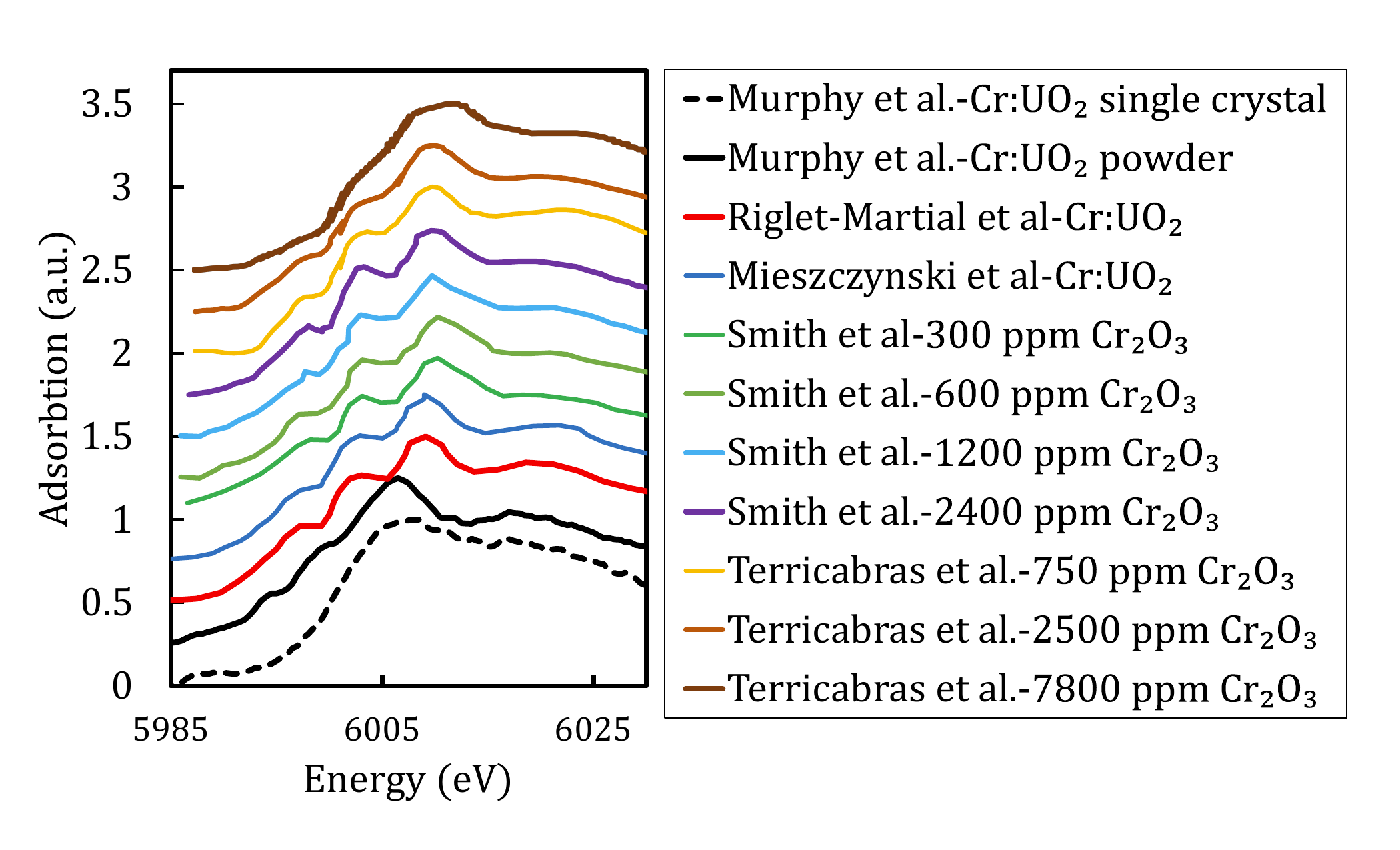}
    \caption{XANES spectra for Cr doped $\ch{UO_2}$ taken from Riglet Martial \cite{RIGLETMARTIAL201463}, Smith \cite{SMITH2022}, Mieszczynki \cite{mieszczynski2014microbeam}, Murphy \cite{MURPHY2023}, and Terricabras \cite{TERRICABRAS2025156114}}
    \label{fig:XANES}
\end{figure}

The lattice solubility is also linked to the charge states of the dopants because the solution energy and preferred solution mode depend on the charge state of Cr as illustrated by the calculations of Cooper \textit{et al} \cite{COOPER2018403}. There has been extensive debate about the charge state of Cr in $\text{UO}_2$. The +3 Cr charge state predicted by the solubility model has been challenged by first principles modeling by Sun \textit{et al.} \cite{SUN2020}. The study found evidence for the lowest energy incorporation mode of Cr into the lattice being in the +2 oxidation state with a charge compensating neighboring O vacancy. The study also reinterprets X-ray adsorption near edge spectroscopy (XANES) spectra by Riglet-Martial \cite{RIGLETMARTIAL201463} and Mieszczynski \cite{mieszczynski2014microbeam} to support +2 oxidation state of Cr, as shown in Figure \ref{fig:XANES}. They proposed that the ionic radius of the Cr +2 state is closer to U +4, reducing the strain. Further XANES measurements by Smith \textit{et al.} \cite{SMITH2022}, provided evidence for Cr +2 substituting for U as well as the presence of a mixed charge state. Moreover, Smith \textit{et al.} showed that multiple Cr +2 environments exist in close proximity in low symmetry motifs, enabling s to p orbital mixing based on the pre-edge features at \SI{5995}{eV}. Murphy \textit{et al.} \cite{MURPHY2023} performed electron paramagnetic resonance (EPR), high energy resolution fluorescence detected (HERDF) XANES, and electron X-ray adsorption fluorescence spectroscopy on a single crystal and powder samples to deconvolute the contributions of Cr from the bulk and grain boundary. The results showed that Cr is in the +3 charge state in the bulk, but at the GB, they found evidence of a mixture of +3/+2/+0 charge states.  XANES spectra for Cr doped $\ch{UO_2}$ are all shown in Figure \ref{fig:XANES}.

Moreover, the EPR results suggest that the charge compensation of the Cr +3 substitutions in the bulk is not achieved by either the oxidation of U to the +5 charge state or an adjacent oxygen vacancy \cite{MURPHY2023}. Instead, they propose that charge neutrality in the bulk is obtained through remote O vacancies. Gascoin \textit{et al} \cite{Gascoin2025} recently used DFT+U to resolve the dependence of Cr charge on the Hubbard U parameters, explaining discrepancies between papers that found the +3  and +2 charge states, respectively. The authors also identified a site dependence of the Cr charge state, with $\ch{Cr^{+1}}$ only soluble in the interstitial site, while the $\ch{Cr^{+2}}$ is the only stable state in a UO divacancy site, and both $\ch{Cr^{+2}}$ and $\ch{Cr^{+3}}$ are stable in the U vacancy, with $\ch{Cr^{+3}}$ being more stable. Looking at Cr twin defects with an oxygen vacancy as proposed by Murphy \textit{et al} \cite{MURPHY2023}, Gascoin \textit{et al.} reproduced the empirical potential findings of Guo \textit{et al.} \cite{guo2017atomic} that the lowest energy configuration was neighboring Cr sharing an oxygen vacancy in contrast to Murphy \textit{et al.}'s findings. They also found that the lowest solution energy configuration was $\ch{Cr^{+2}}$ in the divacancy. The persistence of the discrepancy between theory and experiment points to the computational limitation of first principles calculations from exploring long range interactions with larger supercells and kinetics. Addressing this lack and reconciling experiment with computation will require the next generation of computational methods to account for variable charge states \cite{PANNAv2} and schemes to identify local minima that may arise from treatment of Hubbard U corrections \cite{moore2024high}. Moreover, the findings by Murphy \textit{et al.} also underscore the complexity of the GB phase that will be considered in the next section. 

\begin{figure}
    \centering
    \includegraphics[width=1.0\linewidth]{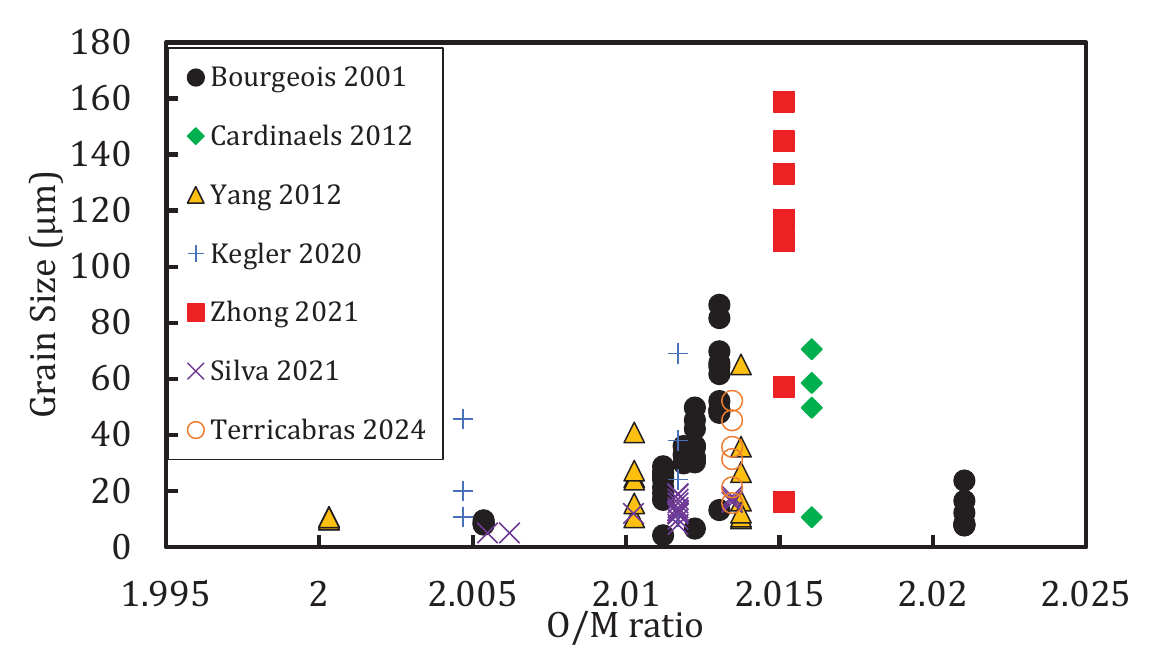}
    \caption{Cr doped $\text{UO}_2$ O/M ratio under reported sintering parameters versus grain size for available literature \cite{BOURGEOIS2001313,cardinaels2012chromia,yang2012effect,kegler2020chromium,zhong2021preparation,silva2021evaluation,TERRICABRAS2024155022} and calculated with equation 35 from reference \cite{watanabe2023oxygen}.}
    \label{fig:O/M ratio versus Grain Size}
\end{figure}

\begin{figure}
    \centering
    \includegraphics[width=1\linewidth]{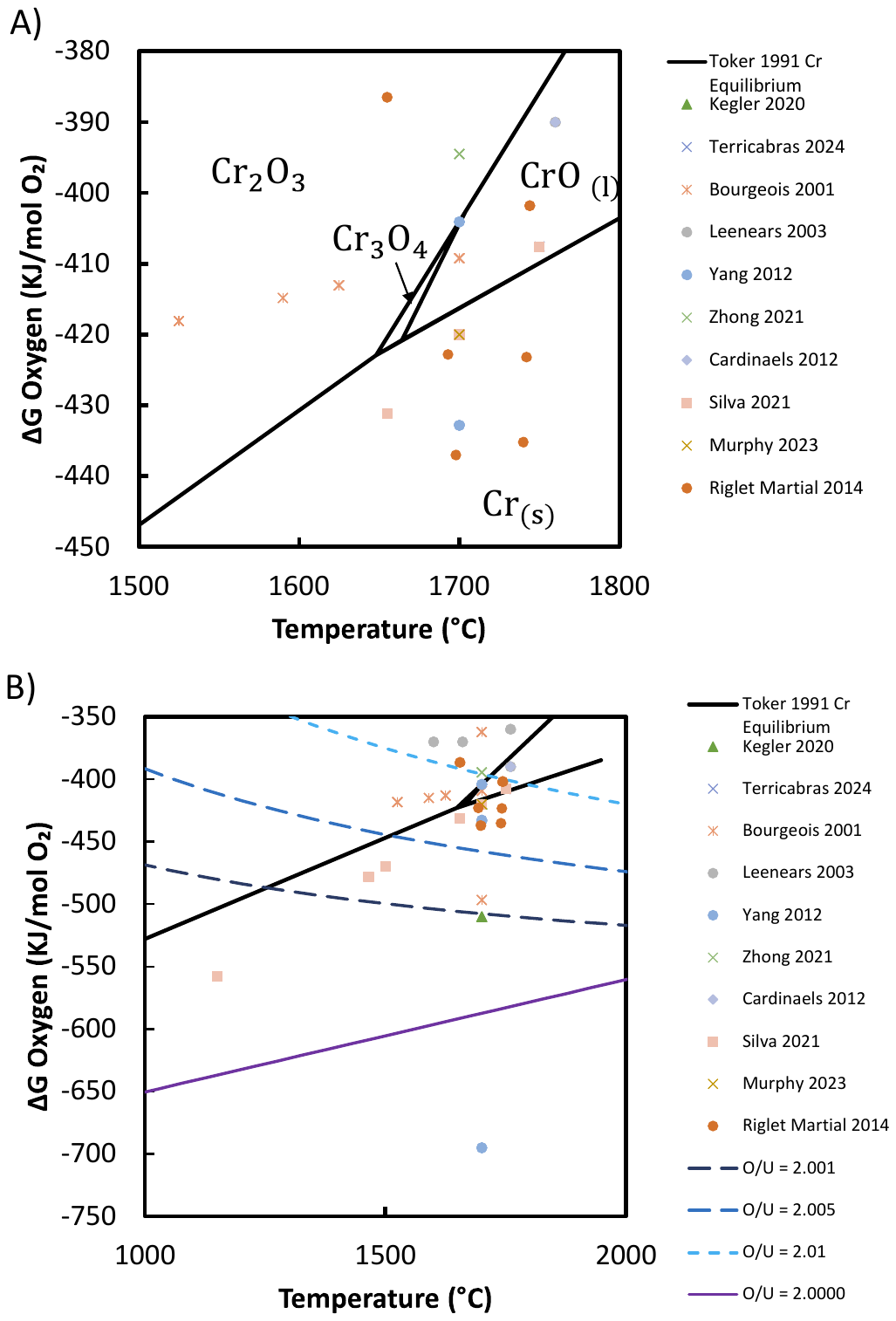}
    \caption{Sintering conditions from previous studies of Cr doped $\text{UO}_2$ \cite{kegler2020chromium,TERRICABRAS2024155022,BOURGEOIS2001313,leenaers2003solubility,yang2012effect,zhong2021preparation,cardinaels2012chromia,silva2021evaluation,MURPHY2023,RIGLETMARTIAL201463} (a) focuses on the boundary between the various Cr-O phases which are annotated from \cite{toker1991equilibrium} (b) shows the full range with the theoretical O/M ratio for $\text{UO}_2$ for the specified conditions calculated from reference \cite{watanabe2023oxygen}.}
    \label{fig:phasediagram}
\end{figure}

\subsection{Cr above its solubility limit}
\begin{figure}
    \centering
    \includegraphics[width=1.0\linewidth]{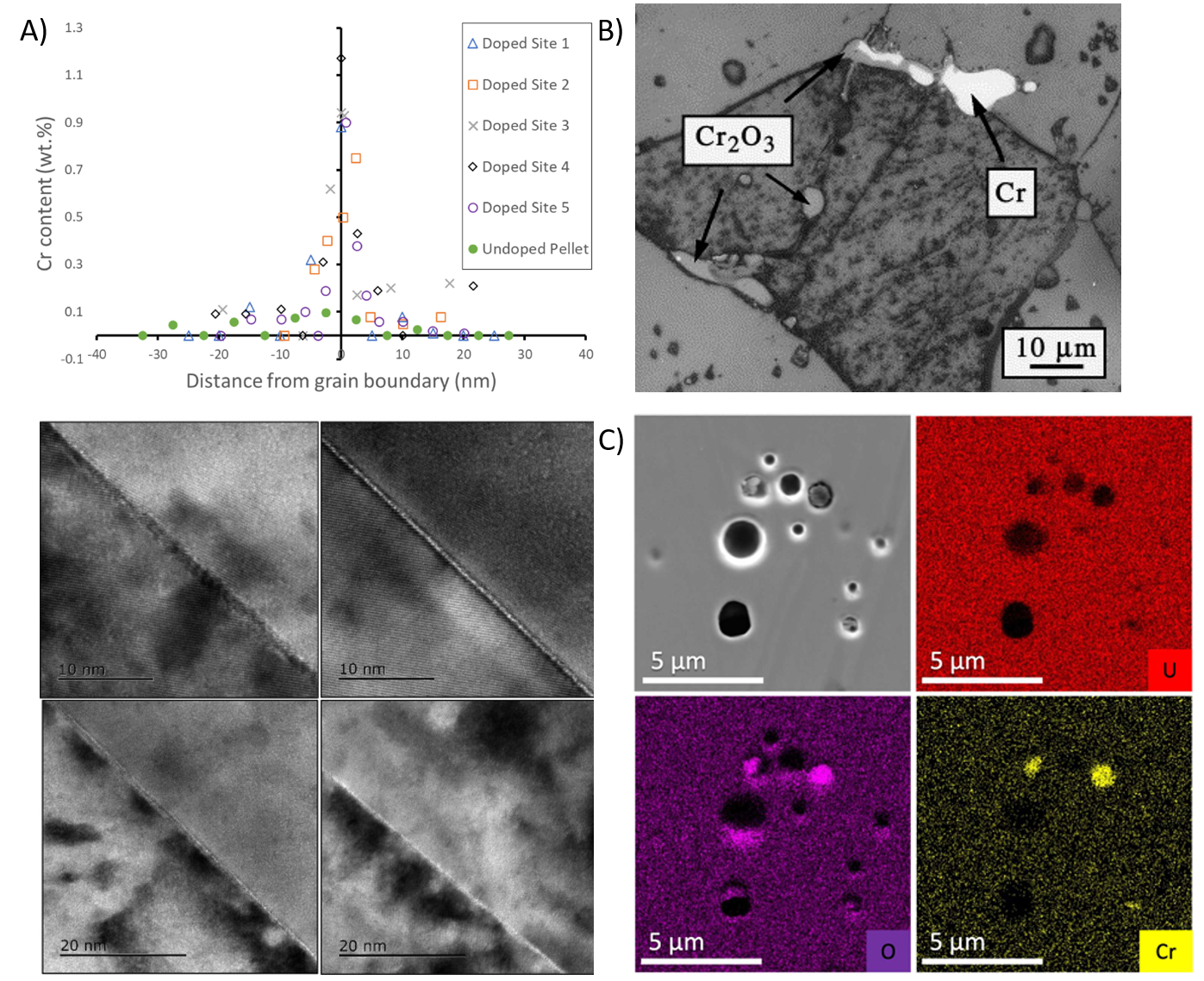}
    \caption{(a) Cr concentration as a function of distance from GB assessed by EDX spectrsocopy and high-resolution transmission electron microscopy images from a Cr-doped $\ch{UO}_2$ pellet, including a grain boundary, taken from Middleburgh \textit{et al.} \cite{MIDDLEBURGH2023154250}, (b) Secondary phase precipitates in $\ch{UO}_2$ pellet initially containing 0.25 wt\% $\ch{Cr_2O_3}$ taken from Bourgeois \textit{et al} \cite{BOURGEOIS2001313}. (c)  SEM image of $\ch{UO}_2$ doped with 0.78 wt\% $\ch{Cr_2O_3}$ with electron dispersive spectroscopy composition mapping of Cr rich precipitate taken from Terricabras \textit{et al.} \cite{TERRICABRAS2024155022}. }
    \label{fig:microscopy}
\end{figure}

As shown by electron micro-probe analysis in Leenaers \textit{et al.} \cite{leenaers2003solubility} and Cardinaels \cite{cardinaels2012chromia} the Cr content added to $\text{UO}_2$ is greater than the measured Cr content in the $\text{UO}_2$ lattice. The loss of Cr is partially due to the volatilization of Cr at high temperature \cite{peres2012high,SMITH2022}) and partially due to heterogeneous regions within the fuel microstructure with higher Cr content, as in Figure \ref{fig:microscopy}. These Cr-enriched regions can be described as a secondary phase within the microstructure that precipitates out of the solid solution. Different types of Cr precipitates can be found in $\text{UO}_2$ based on processing conditions, and they can have a variety of impacts on the fuel performance. 

The distorted local atomic environments of GBs provide a thermodynamic driving force for the enrichment of solutes that can lead to secondary phases in the GB. Bourgeois \textit{et al.} observed Cr rich precipitates at GBs in $\text{UO}_2$ doped with Cr above its solubility limit \cite{BOURGEOIS2001313} see Figure \ref{fig:microscopy} (B). The precipitates Bourgeois \textit{et al.} observed under optical microscopy had a rounded morphology and seemed to consist of a Cr metallic phase and a $\text{Cr}_2\text{O}_3$ oxide phase, suggesting that these precipitates formed from a liquid eutectic phase during cooling from the sintering temperatures. Meanwhile, Leenaers \textit{et al.} \cite{leenaers2003solubility} and Cardinaels \textit{et al.} \cite{cardinaels2012chromia} both identified $\text{Cr}_2\text{O}_3$ precipitates of about 3$\mu{m}$ size in Cr-doped $\text{UO}_2$ based on EPMA results. The authors also found one precipitate with CrO stoichiometry for the sample sintered at an oxygen potential of \SI{-390}{kJ/mol} $\text{O}_2$ and temperature of \SI{1760}{\celsius}, which falls in the region of CrO stability predicted by Toker \cite{toker1991equilibrium}. These results are consistent with the findings of Riglet-Martial \textit{et al.} that showed the stoichiometry of Cr precipitates depended on the thermodynamically stable Cr-O phase with metallic Cr and $\text{Cr}_2\text{O}_3$ precipitates in the corresponding thermodynamic regions shown in Figure \ref{fig:phasediagram} (a) \cite{RIGLETMARTIAL201463}. They found that for sintering conditions where CrO was stable, precipitates had stoichiometries close to CrO, but they disassociated into bi-phasic regions like those seen in Bourgeois \textit{et al}'s study. Devillaire \textit{et al.} \cite{devillaire2023characterisation} also found precipitates with stoichiometry of 8/7 Cr to O, and suggest a U stabilized $\text{Cr}_3\text{O}_4$ phase supports the findings of Riglet-Martial \textit{et al}.

Metallic Cr precipitates have been found to be detrimental to grain growth in doped $\text{UO}_2$ as they lead to GB pinning because U and O cannot diffuse through them and the precipitates create an energetic barrier to GB migration \cite{yang2012effect,peres2012high}. In $\text{UO}_2$ doped with Cr above its solubility limit, $\text{Cr}_2\text{O}_3$ precipitates were identified with XRD by Kuri \textit{et al.} \cite{KURI2014158}. These precipitates had a corundum crystal structure similar to chromia, but with added structural disorder likely explained by the interface effects between the precipitates and the $\text{UO}_2$ lattice. $\text{Cr}_2\text{O}_3$ precipitates are found both in the grain interior as in Figure \ref{fig:microscopy} C) and at GBs \cite{SMITH2022, TERRICABRAS2024155022} where they can inhibit the inter-granular diffusion of ions at the onset of sintering \cite{peres2012high}. Kegler \textit{et al.} found no preferred precipitation of secondary phases within the microstructure of Cr doped fuel prepared by wet chemical approaches, illustrating the dependence of Cr precipitate behavior on processing parameters \cite{kegler2020chromium}.  $\text{Cr}_2\text{O}_3$ precipitates have even been found in samples doped below the Cr solubility limit, though this might point to incomplete homogenization of the powders during mechanical ball milling \cite{milena2021raman}. However, Middleburgh \textit{et al.} have provided direct evidence via TEM of the enrichment of Cr at $\ch{UO_2}$ GBs and measured via EDX linescan as shown in Figure \ref{fig:microscopy} (A) \cite{MIDDLEBURGH2023154250}. A follow up combined experiment and modeling study found good agreement between enhanced Cr concentrations at GBs measured with EDS (7000 wppm) with Cr concentrations calculated using segregation energies obtained with empirical interatomic potentials (25,000-100 wppm) \cite{TERRICABRAS2025156114}. Overall, Cr-doped $\text{UO}_2$ can be considered as a composite between a $\text{UO}_2$ matrix with some Cr substituting for U, with Cr rich precipitates whose stoichiometry is determined by the redox reactions between metallic Cr and oxide phases. 


\section{Properties and performance of Cr-doped uranium dioxide}

\subsection{Fission gas release and swelling}\label{FGR}
While FGR theoretically scales inversely with grain size, early studies found a more complex relationship \cite{KILLEEN1980177}. In principle, Cr doping can also change the diffusivity of fission gas. Higher fission gas diffusivity increases the rate of FGR, thereby competing with the larger grain size. In a study by Killeen, irradiated $\text{UO}_2$ doped with 5000 wppm $\text{Cr}_2\text{O}_3$ had similar gas release and swelling to a reference $\text{UO}_2$ fuel \cite{KILLEEN197539}. This result is despite the doped sample having larger grain sizes of $\SI{70}{\mu m}$ versus $\SI{10.2}{\mu m}$ for undoped $\text{UO}_2$  samples \cite{KILLEEN1980177}. Killeen argues that the similar FGR must be due to enhanced fission gas diffusion rates. The finding is further supported by Kashibe and Une, who report a 3 times higher diffusion coefficient for Xe in $\text{UO}_2$ doped with 650 wppm $\text{Cr}_2\text{O}_3$ after irradiation \cite{kashibe1998effect}. Arborelius \textit{et al.} in 2006 report a lower FGR in Cr doped $\text{UO}_2$, with 50\% less FGR than the standard $\text{UO}_2$ sample during post irradiation examinations \cite{arborelius2006advanced}. The samples used by Arborelius were doped with 1000 ppm $\text{Cr}_2\text{O}_3$ had a 3D grain size of $\SI{44}{\mu m}$ and a FGR of 17.2\% compared to the reference $\text{UO}_2$ with a 3D grain size of $10-\SI{12}{\mu m}$ and FGR of 30.2\%\footnote{Arborelius \textit{et al.} used 3D grain size defined as the linear intercept with a correction factor of 1.5 instead of the traditional grain size measurement making it hard to compare with other studies in the literature.}. However this 50\% reduction is conflated because the Cr doped sample was held at the maximum power for \SI{7.7}{h} instead of the \SI{12}{h} of the undoped sample. In a different experiment, a sample co-doped with 500 ppm $\text{Cr}_2\text{O}_3$ and 200 ppm $\text{Al}_2\text{O}_3$ with a grain size of $\SI{52}{\mu m}$ showed 20.5\% FGR compared to 29.7 \% FGR from an undoped test sample. The experiments by Killeen were performed at low burnup, whereas Arborelius \textit{et al.} showed larger FGR for 30 MWd/kgHM due to differences in testing methodology, as their specimens were punctured to measure the FGR in contrast to in  reactor studies. 

Although Arborelius opined that larger grain size outweighs the effect of doping on enhancing fission gas diffusivity, Killeen's samples, despite exhibiting larger grain size, show no corresponding reduction in FGR. This suggests underlying differences in diffusivity. Direct comparison between studies is challenging because diffusivity is sensitive to experimental parameters such as sintering temperature. For instance, Killeen used a $\text{Cr}_2\text{O}_3$ concentration five times greater than that used by Arborelius \textit{et al.}, pointing to the possible dependence of fission gas diffusivity on Cr concentration. Molecular dynamics studies support an increase in self diffusivity with Cr concentration in $\ch{UO_2}$, but factors such as the high dopant concentrations and amorphous structures in that study prevent broad generalization of the results \cite{OWEN2023154270}. Che \textit{et al.} \cite{CHE2018271} first modeled Cr doped $\text{UO}_2$ with a sensitivity analysis on the BISON fuel performance code \cite{HALES2013531}. Observing a low Cr solubility limit at reactor operation temperature of ($<$\SI{1700}{K}), while Cooper \textit{et al} \cite{COOPER2021152590} developed a mechanistic model of fission gas diffusivity for Cr doped $\text{UO}_2$ based on the defect cluster dynamics model of Matthews \textit{et al.} \cite{MATTHEWS2020152326} and Gamble \textit{et al.} \cite{gamble2019atf} that could be integrated into the BISON fuel performance code. They showed how the Cr/$\text{Cr}_2\text{O}_3$ equilibrium in Figure \ref{fig:phasediagram} controls the oxygen potential, which in turn alters the concentration of defect clusters responsible for fission gas diffusion. They benchmark their code against the IFA-716 irradiation experiments of doped fuels in the Halden reactor \cite{tverberg2014update,bremont2011ifa}. Cooper \textit{et al.} argued that Cr doping raises the oxygen potential at high temperature and lowers it at low temperature, promoting the concentration of the defect clusters responsible for Xe diffusion by increasing the O/M ratio. The defect cluster model predicts an increase in the diffusivity by a factor of three at reactor operation conditions, which fits the experimental trend from Kashibe \cite{kashibe1998effect}. Further work has taken the physics informed model for  Cr-doped $\text{UO}_2$ and applied a variational Bayesian Monte Carlo method to tune the BISON parameters to capture the effect of co-doping with $\text{Al}_2\text{O}_3$\cite{che2021application}. The application of machine learning methods and uncertainty quantification to the doped $\text{UO}_2$ is promising and will benefit from additional modeling and experimental data. Moreover, in recent years RAVEN analyses have been performed over BISON input properties that are expected to impact fuel performance for doped fuels \cite{cheniour2023sensitivity}. Additionally, this BISON Model is being used to design separate effects experiments for validation of the model \cite{gorton2024modeling}. Nicodemo \textit{et al.} adapted the model based on oxygen potential to consider the effect of the experimentally reported Cr solubility into the $\text{UO}_2$ lattice on the defect equilibria and pointed to the fact that Cr incorporation as an interstitial promotes the concentration of U vacancies and hence the diffusivity of fission gas \cite{nicodemo2024chromia}. While Nicodemo \textit{et al.} consideration of the experimental Cr solubility seems rigorous, they misused the defect mechanism of Murphy \textit{et al} \cite{MURPHY2023}, which asserts that Cr incorporates substitutionally for U and would hence lower the concentration of U vacancies, instead of promoting U vacancies through Schottky and Frenkel equilibria, which can only occur in the interstitial case. If the defect mechanism of Murphy \textit{et al} is used to produce the concentration of O vacancies, $[\ch{v^{**}_O}]=1/2[\text{Cr}_\text{U}']$, then combining the Schottky equilibrium condition $ [\ch{v^{**}_O}]^2[\ch{v_U''''}]=K_s $, yields $[\ch{v_U''''}]=4K_s/[\text{Cr}_\text{U}']^2$ instead of the predicted increase in U vacancies. In fact, competition between the oxygen potential effects and Cr substitutional defect chemistry could account for the slight overestimation of experimental Xe diffusivity in the models of Cooper \textit{et al} \cite{COOPER2021152590}. Modeling efforts are poised to improve their predictions of fission gas diffusivity as they consider both the macroscopic effects of Cr/$\text{Cr}_2\text{O}_3$ redox and the microscopic effects of Cr on the $\text{UO}_2$ lattice, which can in turn be fed into larger length scale fuel performance codes. Modeling methods are well situated to quantify the effect of Cr doping on fission gas diffusivity under various reactor operation and accident conditions so that the increase in diffusivity can be out-competed by the larger grain sizes promoted by Cr doping, leading to reduced FGR in Cr-doped fuels. 

The effect of Cr doping on the irradiation behavior of $\text{UO}_2$ with respect to swelling is limited and only partially understood. Killeen found that Cr-doping led to no change in the fission gas swelling of fuel specimens \cite{KILLEEN1980177}. Arborelius \textit{et al} also found a similar swelling rate between their doped and undoped specimens and 0.1\% higher axial rod growth for the undoped specimens \cite{arborelius2006advanced}. However, in a follow up study by Westinghouse, they report the same specimens had larger volumetric expansion when irradiated due to starting at a higher density \cite{backman2010westinghouse}. Because the steady-state rate of fission gas swelling is consistent between doped and undoped fuels, it is considered a worthwhile tradeoff for enhanced fission gas inventory, improved viscoplasticity, and oxidation resistance. While Fraczkiewicz found via Transmission Electron Microscopy of ion-irradiated samples that chromium makes the coalescence of irradiation defects easier by stabilization of defect clusters due to precipitation of Cr \cite{fraczkiewicz2010study}, the significance of this interaction does not appear in measures of reactor performance \cite{backman2010westinghouse}. Mieszczynski \textit{et al.} performed XRD on doped and undoped fuel samples before and after irradiation and found that the expansion of the lattice parameter due to irradiation massively outweighed the small difference in lattice parameter between the pristine doped and undoped fuels \cite{mieszczynski2014irradiation}. Further irradiation studies are planned to fill in the gaps in the study of how Cr doping affects the irradiation response of $\text{UO}_2$ \cite{gorton2024modeling}. 

\subsection{Thermophysical properties - Thermal conductivity, heat capacity and thermal expansion}
\begin{figure}
    \centering
    \includegraphics[width=1.0\linewidth]{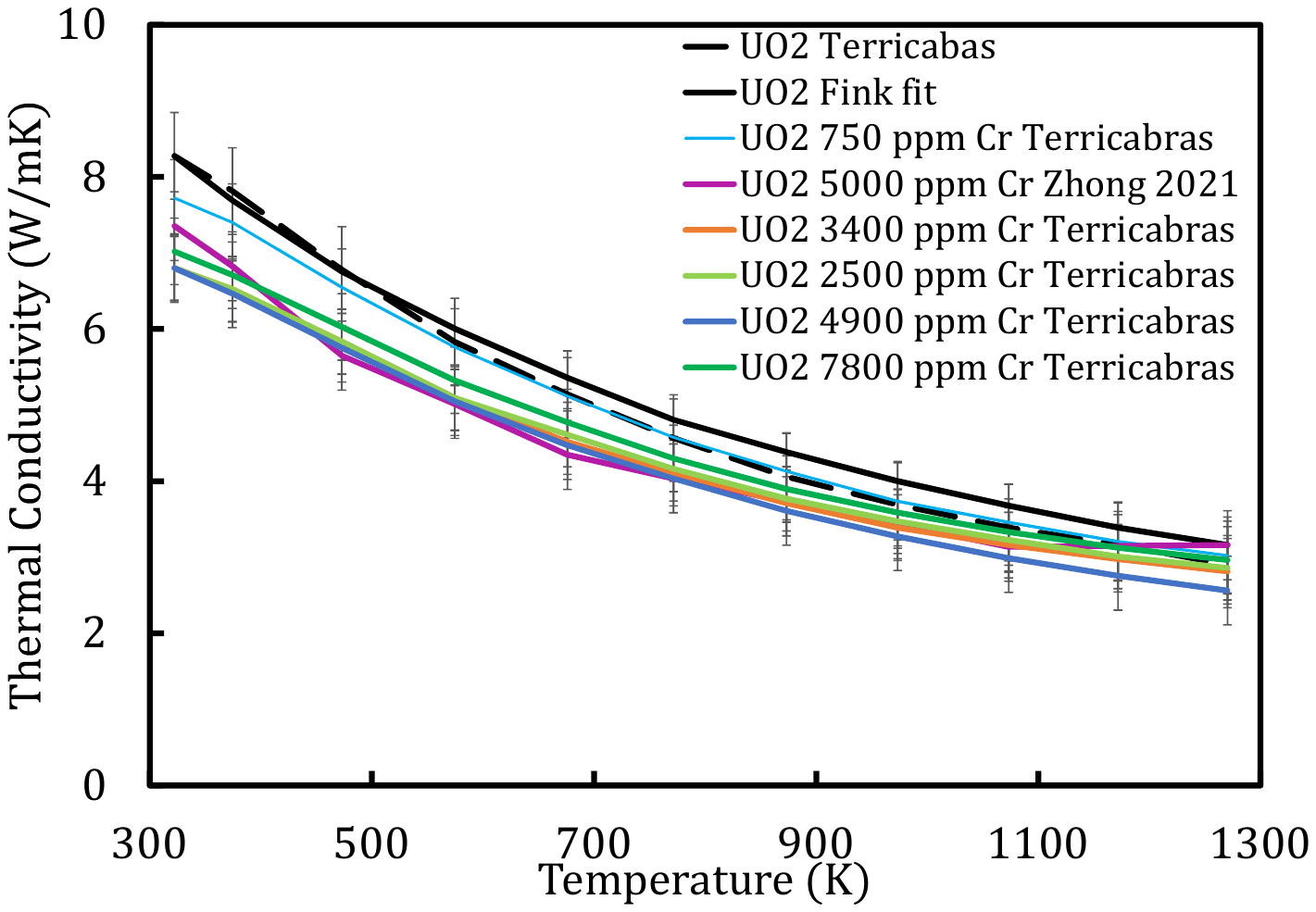}
    \caption{Thermal conductivity versus temperature for Cr doped $\text{UO}_2$ at different levels taken from data in Terricabras \textit{et al} \cite{TERRICABRAS2024155022}, Zhong \textit{et al.} \cite{zhong2021preparation}, and fit for $\text{UO}_2$ from Fink \cite{fink_thermophysical_2000}}
    \label{fig:Thermal Conductivity}
\end{figure}

Arborelius \textit{et al.} \cite{arborelius2006advanced} reported that the thermophysical properties which includes the heat capacity, thermal expansion coefficient, melting temperature, and thermal diffusivity showed no measurable difference between the doped and undoped samples. This trend is consistent with the values reported by Fink \cite{FINK20001}. While there is no significant improvement in the thermal conductivity of the large grained Cr doped samples, a thermal shock test showed fewer circumferential cracks, which were attributed to better heat transfer capability. Zhong \textit{et al.} report similar thermal conductivities/diffusivity for undoped and Cr doped specimens as shown in Figure \ref{fig:Thermal Conductivity} with the Cr doped samples having slightly lower thermal conductivity below \SI{400}{\celsius} and higher thermal conductivity above \SI{800}{\celsius} which they attributed to the effects of impurities and the reduced effect of grain boundary phonon scattering for high grain boundaries \cite{zhong2021preparation}. Terricabas \textit{et al.} also did not find significant changes in the specific heat capacity and coefficient of thermal expansion of doped $\text{UO}_2$, but they found a decrease in the thermal conductivity for Cr doping above 750 ppm $\text{Cr}_2\text{O}_3$ by 15\% as seen in Figure \ref{fig:Thermal Conductivity} \cite{TERRICABRAS2024155022}. Interestingly, larger grain sizes do not lead to higher thermal conductivity as might be due to lower phonon scattering at GBs, instead dopants appear to promote lattice phonon scattering. Also, the reduction in thermal conductivity reported in Terricabas \textit{et al.} and reproduced in Figure \ref{fig:Thermal Conductivity} shows thresholding behavior where the lower thermal conductivity does not vary with doping levels above 750 ppm $\text{Cr}_2\text{O}_3$. This suggests a relationship between the Cr solubility limit and the decreased thermal conductivity but requires further systematic investigations. At the low doping concentrations used in industry, the thermophysical properties of Cr doped $\text{UO}_2$ closely approximate those of $\text{UO}_2$. The work by Terricabas \textit{et al.} is valuable for assessing the role of doping level on the thermophysical properties of unirradiated fuel, but additional work is necessary to understand the change in thermal conductivity of doped fuel under irradiation conditions. 

Arborelius \textit{et al} did not find significant differences in the heat capacity and coefficients of thermal expansion between their Cr-doped and undoped specimens \cite{arborelius2006advanced}. Heat capacity is not sensitive to small additions of dopants $<$0.2 wt\% \cite{massih2014effects}.  Terricabas \textit{et al.} also report no statistically significant differences in the heat capacity or coefficient of thermal expansion for their various levels of doping within the precision of their measurement technique \cite{TERRICABRAS2024155022}. Concentrations of dopants that leave the $\ch{UO_2}$ lattice intact would not be predicted to have large impacts on the heat capacity. Additionally, the lattice strain due to dopants is small, so the effect of doping on thermal expansion would be expected to be likewise small in magnitude, as supported by the current experimental findings. 

\subsection{Chemical properties and pellet cladding interactions}
Beyond the effect of Cr doping on the physical properties of the fuel in the reactor, it is important to understand the effect of Cr doping on the chemical reactivity of $\text{UO}_2$. The interaction with cladding is also another important aspect that must be considered, especially as the diffusion of Cr towards the cladding may alter the mechanical \cite{delafoy2006areva,delafoy2018benefits} and chemical interactions with the cladding materials \cite{cui2023thermodynamic, konarski2019thermo}. Arborelius \textit{et al.} reported that the oxidation rate of the Cr doped pellets is half that of the undoped pellets based on a thermal microbalance test at \SI{400}{\celsius} \cite{arborelius2006advanced}. This result is supported Milena-Perez \textit{et al.}, who looked at Cr doped $\text{UO}_2$ before \cite{MILENAPEREZ2023154502} and after \cite{milena2021raman} irradiation with Raman spectroscopy. In their results for fuels with different doping levels prior to irradiation, there was no difference between the rate of oxidation of Cr-doped and undoped $\text{UO}_2$ in air, but at 1\% oxygen, Cr doping slowed down the oxidation rate, shifting it to a higher temperature and increasing the amount of unoxidized $\text{UO}_2$ and partially-oxidized $\text{U}_3\text{O}_7$ , while reducing the amount of $\text{U}_3\text{O}_8$ detected by Raman spectroscopy. Milena-Perez \textit{et al.} attribute this behavior not to the thermodynamic stabilization of the $\text{UO}_2$ phase by the Cr, but by the slower oxidation kinetics associated with the larger grain size. The lower oxidation rate of Cr-doped $\text{UO}_2$ appears primarily as an effect of increased grain size, reducing the oxidation kinetics. 

Several studies on Cr doped $\text{UO}_2$ have been carried out recently to understand the dissolution rate in varying corrosive environments. For instance, experimental studies of the leaching of the fuel matrix into basic aqueous environments show a decrease in the concentration of dissolved uranium for Cr-doped fuel \cite{cachoir2021evolution}. Nilsson \textit{et al.} report  that the decreased release of fission products into aqueous media is attributed to the larger grain sizes having slower dissolution kinetics, as there is less surface area for dissolution \cite{NILSSON2017123}. When the dissolution rate is normalized by the surface area, the dissolution rate for undoped and Cr-doped $\text{UO}_2$ are within the experimental uncertainties \cite{barreiro2021aqueous}. However, Smith \textit{et al.} looked at specimens of varying Cr concentration with consistent grain size produced via sintering in a reducing environment and found that Cr doping slows the normalized dissolution rate of the $\text{UO}_2$ matrix near room temperature \cite{smith2023oxidative}. Smith \textit{et al.} attribute this to the thermodynamically favorable galvanic coupling between $\text{U}^{+6}$ and $\text{Cr}^{+2}$, though the debate on the Cr charge state is the subject of section \ref{solubility}. The dissolution rate in the bicarbonate medium is proportional to the concentration of Cr when the temperature is raised to \SI{60}{\celsius}. This result suggests that Cr and its precipitates influence the surface reactivity. For perspective, Cr doping has much less of an impact on the dissolution rate of the fuel in geological storage than the aqueous medium the fuel is in  contact with during reactor operations \cite{RODRIGUEZVILLAGRA2022153880,rodriguez2025impact}. Therefore, while Cr doping might have beneficial or mixed effects on the dissolution of the fuel, the evidence suggests that the differences in the dissolution rates of Cr doped fuel and undoped fuels are not significant and may be attributed to the beneficial effects of increased grain size. 

The large grain sizes obtained from doping are also desirable for improved pellet cladding chemical/ mechanical interaction. Arborelius reports Cr doped $\text{UO}_2$ experiencing less cracking and fewer incipient cracks after thermal shock and power ramp/bump tests \cite{arborelius2006advanced}. Conversely, Lupercio \textit{et al.} report a 20\% decrease in the fracture stress for 3000 wppm $\text{Cr}_2\text{O}_3$ doped $\text{UO}_2$ under ball-on-ring equibiaxial flexure testing suggesting doping promotes crack formation \cite{lupercio2025statistical}. Lupercio \textit{et al.} found a trans-granular fracture mode facilitated by Cr-O precipitates and residual lattice stresses; these trans-granular fractures would accelerate the release of fission gas that is trapped on the lattice.  The reduction in fracture stress is desirable for the production of many small cracks in the periphery of the fuel, but this work is under normal operation, which means the first ramp to power, and the behavior relevant to PCI will be altered by the presence of fission products and radiation damage. While there is theoretical work from Curti \textit{et al.} via a thermodynamic analysis with a sublattice model that found other dopants like Mo influence the oxygen chemical potential, Cr in the cladding did not impact the O potential \cite{CURTI2020152140}. 

\section{Mechanisms for Cr-doped accelerated grain growth}
\subsection{Experimental observations of enhanced grain growth}
\begin{table}[ht]
\noindent
\caption{Optimal sintering parameters from literature for largest reported grain size of $\ch{Cr_2O_3}$ doped $\ch{UO_2}$}
\label{tab:sintering_params}
\renewcommand{\arraystretch}{1.5}
\begin{tabularx}{450pt}{llllllllX}
\hline
Grain size & $\text{Cr}_2\text{O}_3$& $T_{s}$  &  Atmosphere      & $\mu_O$ & $a_{\ch{UO_2}}$ & O/M  & Ref.  & Note      \\
($\mu\SI{}{m}$)  & (wppm) & (\SI{}{\celsius})  &       & ($\SI{}{kJ/mol}$) & (\SI{}{\angstrom}) & ratio & &       \\
\hline
17 & 1000& 1675 & $\ch{N_2}$-4.7\% $\ch{H_2}$  & - & 5.47 & - & \cite{milena2021raman} & Samples for Raman spectroscopy \\
69  & 2500 & 1700 & Ar/$\ch{H_2}$  & -420 & 5.46 & 2.012  & \cite{kegler2020chromium} & Optimizing preparation technique  \\
52 & 4883 & 1750 & Ar-4\%$\ch{H_2}$  & -408 & 5.47 & 2.014 & \cite{TERRICABRAS2024155022} & Thermophysical properties measurements\\
13 & 106 & 1700 & $\ch{N_2}$-5\% $\ch{H_2}$  & - & 5.47  & - & \cite{SMITH2022} & Reducing environment small grain  \\
86 & 2494  & 1700& $\ch{H_2}$/$\ch{H_2O}$ 1\%& -409 & - & 2.013 & \cite{BOURGEOIS2001313}  & Optimized wt\% Cr, T, and $\mu_O$ \\
100  & 1930  & 1700  & 0-1.7\% $\ch{CO_2}$ & -404 & - & -& \cite{yang2012effect} & Stepwise $\mu_O$ to control Cr at GB \\
159 & 4997& 1700 & $\ch{CO_2}$/$\ch{H_2}$ 2\%& -394& - & 2.015 & \cite{zhong2021preparation}  & Large grain size via wet preparation\\
71  & 1578 & 1760 & $\ch{H_2}$- 2.7\%$\ch{H_2O}$  & -390  & 5.47   & 2.016    & \cite{cardinaels2012chromia} &  Investigation of lattice parameter \\
44 & 1000 & 1800  & $\ch{H_2}$/$\ch{CO_2}$  & - &- & -& \cite{arborelius2006advanced} & Westinghouse ADOPT Fuel \\
19 & 1222 & 1700 & Ar-4\%$\ch{H_2}$ & -420 & 5.46  & 2.012 & \cite{silva2021evaluation}  & Effect of increasing dopant concentration of grain growth during sintering\\
15 & 650 & 1750 & $\ch{H_2}$ & - & - & - & \cite{kashibe1998effect} & Reducing environment small grain  \\ 
\hline
\end{tabularx}
\end{table}

Previous work has explored the sintering of Cr doped $\text{UO}_2$ under different temperature and oxygen chemical potential conditions near the domains of co-stability of Cr metal, CrO, and $\text{Cr}_2\text{O}_3$ as shown in Figure \ref{fig:phasediagram}. The thermodynamic range where $\text{Cr}_3\text{O}_4$ is stable is quite narrow and difficult to obtain experimentally, though some work has found $\text{Cr}_3\text{O}_4$ precipitates at the GBs of Cr doped $\text{UO}_2$ \cite{devillaire2023characterisation}. This points out that the thermodynamic stability of Cr-O differs from the bulk because of the surrounding $\text{UO}_2$ matrix. 

The foundational work for understanding why Cr doping of $\text{UO}_2$ leads to accelerated grain growth is Bourgeois \textit{et al} \cite{BOURGEOIS2001313}. Sintering $\text{UO}_2$ in $\text{H}_2$ atmosphere with controlled concentrations of 0.05, 1, and 5 vol.\% $\text{H}_2\text{O}$ in $\text{H}_2$, Bourgeois \textit{et al} report that the largest grain sizes are obtained only at the 1 vol.\% $\text{H}_2\text{O}$ with two peaks in mean grain size versus Cr concentration at 700 ppm and 2500 ppm $\text{Cr}_2\text{O}_3$. Moreover, the grain size is also proportional to the temperature within the range of \SI{1525}{\celsius} to \SI{1700}{\celsius}. The authors attribute the second peak in grain size above their estimated solubility limit of 700 ppm to a Cr-$\text{Cr}_2\text{O}_3$ eutectic. The eutectic accelerates grain growth through liquid phase sintering above \SI{1550}{\celsius} but below the melting temperature of $\text{Cr}_2\text{O}_3$. A Cr-$\text{Cr}_2\text{O}_3$ eutectic with CrO composition is supported by thermodynamic work by Toker \cite{toker1991equilibrium}. The work of Bourgeois \textit{et al} \cite{BOURGEOIS2001313} provides evidence for the claim that the CrO phase is responsible for the enhanced grain growth as the 1 vol.\% $\text{H}_2\text{O}$ ($\mu_{O_2}=-\SI{410}{kJ/mol}$ at \SI{1700}{\celsius} falls within the region of CrO stability whereas 0.05 and 5 vol.\% $\text{H}_2\text{O}$ are in the regions of Cr and $\text{Cr}_2\text{O}_3$ stability. However, below \SI{1700}{\celsius}, the bulk stable phase reverts to $\text{Cr}_2\text{O}_3$ phase for the oxygen potentials used in the study, pointing to the eutectic and the complexities introduced by interaction with$\text{UO}_2$. This study has guided the further exploration of the temperature and oxygen potential space relevant to the sintering of Cr doped $\text{UO}_2$. 

Arborelius \textit{et al.} do not comment on the mechanism behind grain growth beyond observing that Cr concentration is the main driver of larger grain sizes at concentrations predicted suboptimal by Bourgeois \textit{et al.} though without reporting their sintering atmosphere. Yang \textit{et al.} pioneered a method to increase the oxygen potential in a stepwise manner during isothermal sintering to enhance grain growth of Cr-doped $\text{UO}_2$ \cite{yang2012effect}. They observe that Cr dissolution requires the disappearance of metallic Cr precipitates, with the segregation of $\text{Cr}^{+3}$ ions to the GB being the proposed mechanism for grain growth linked to the variation in the oxygen potential, as it controls the Cr solubility. Cr is responsible for increased self diffusivity. The doped sample produced undulating GBs with a larger than necessary surface area, pointing to either lowering the GB energy or Zener drag, which could be caused by Cr doping. Zhong \textit{et al.} considered the impact of the mode of Cr incorporation and found that metallic Cr powder had less of an effect on grain growth during sintering than the $\text{Cr}_2\text{O}_3$ powder instead of metallic Cr powder \cite{zhong2021preparation}. It should be noted that the maximum reported grain sizes of Zhong \textit{et al} ($\SI{158}{\mu m}$) 50\% larger than the next highest reported by Yang \textit{et al.} ($\SI{99}{\mu m}$) as shown in Figure \ref{fig:O/M ratio versus Grain Size} and Table \ref{tab:sintering_params}, pointing to the importance of well optimized processing parameters to obtaining maximal grain size. All of these results center attention around the region of CrO stability associated with a liquid sintering mechanism. 

One processing parameter is the forming gas used to obtain the target oxygen chemical potential as shown in Table \ref{tab:sintering_params}. The $\ch{Cr_2O_3}$ vapor pressure increases with how reducing the atmosphere is \cite{TERRICABRAS2024155022} making it challenging to fabricate doped $\ch{UO_2}$ with a known Cr content. Studies reporting post sintering grain sizes above $\SI{20}{\mu m}$ sintered in a gas environment of $\text{H}_2\text{O}$ in $\text{H}_2$ \cite{BOURGEOIS2001313, cardinaels2012chromia} or other wet sintering environments $\text{H}_2$ in Ar \cite{TERRICABRAS2024155022, kegler2020chromium} or $\text{CO}_2$ in $\text{H}_2$ \cite{arborelius2006advanced,yang2012effect,zhong2021preparation}. Reported grain growth in dry $\text{H}_2$ in $\text{N}_2$ \cite{milena2021raman, smith3assessment} and $\text{H}_2$ in Ar \cite{silva2021evaluation} , have been less pronounced than under wet sintering conditions. These results taken together point to the role of oxidative sintering conditions in promoting large grained $\text{UO}_2$. Based on the defect thermodynamics of $\text{UO}_2$, oxidative sintering conditions promote the concentration of U vacancies essential to grain boundary mobility \cite{NIELSON2024} playing a synergistic role with Cr doping. 

The concentration of $\text{Cr}_2\text{O}_3$ powder added is another important processing parameter. Bourgeois \textit{et al} attribute their double peaks in $\text{Cr}_2\text{O}_3$ concentration at 750 ppm and 2500 ppm to the effect of solute drag \cite{BOURGEOIS2001313}. Yang \textit{et al.} \cite{yang2012effect} and Zhong \textit{et al.} \cite{zhong2021preparation} sintering in $\text{CO}_2$ in $\text{H}_2$ found a direct relationship between the $\text{Cr}_2\text{O}_3$ concentration and the grain size. Zhong \textit{et al.} \cite{zhong2021preparation} and Terricabas \textit{et al.} \cite{TERRICABRAS2024155022} reporting their largest grain size of $\text{UO}_2$ at 5000 ppm $\text{Cr}_2\text{O}_3$ double that of Bourgeois \textit{et al}. While the difference in optimal Cr concentration for maximizing grain size might depend on the wet atmosphere used which includes $\text{CO}_2$ in $\text{H}_2$, $\text{H}_2\text{O}$ in $\text{H}_2$, $\text{H}_2$ in Ar. It is also important to note that the reported $\text{Cr}_2\text{O}_3$ added initially differs from the dissolved Cr present to influence grain growth due to the volatilization of Cr under sintering conditions, with Terricabas \textit{et al.} only finding a concentration equivalent to 1500 ppm $\text{Cr}_2\text{O}_3$ in their nominal 4882 ppm sample \cite{TERRICABRAS2024155022, SMITH2022}. This disconnect provides a critical caveat in the existing data and the importance of further work to map out and control the actual Cr concentration rather than the nominal concentration.

Under dry sintering environments, the relationship between grain size and Cr content is clearer with the maximum grain size obtained around 1000-1200 ppm $\text{Cr}_2\text{O}_3$ in the studies of  Silva \textit{et al.} and Milena-Perez \textit{et al.} \cite{silva2021evaluation, milena2021raman}. However, Smith \textit{et al.} found that Cr doping decreased the grain size compared to the undoped reference \cite{smith2023oxidative}. This would seem to point to the expected behavior of solute pinning of the GB during which GB mobility is limited.  Overall, the grain growth behavior of Cr doped $\text{UO}_2$ differs between wet and dry sintering environments, with wet environments producing larger grain sizes at higher Cr concentrations, whereas dry environments produce smaller grain sizes and show less Cr solubility. Empirically, to achieve large grained $\text{UO}_2$ under Cr doping, the standard recipe is to sinter $\text{UO}_2$ doped with enough $\text{Cr}_2\text{O}_3$ to obtain a concentration after sintering equivalent to 1500 ppm at an oxygen potential around -\SI{410}{kJ/mol} and a temperature above \SI{1700}{\celsius} in a wet environment for about 8-10 hours. 

Other low temperature sintering techniques have been attempted, including Field-assisted Sintering techniques (FLASH) sintering \cite{kardoulaki2019report} and Hot isobaric pressing (HIP) \cite{cordara2020hot}; however, these methods have not shown the same increase in grain size of doping as they are several hundred degrees below the relevant thermodynamic regimes. Similar eutetics formed during conventional sintering may be absent due to the lower temperature. 

\subsection{Mechanistic explanations of enhanced grain growth}
There is evidence for multiple effects of Cr doping on the grain growth rate of $\text{UO}_2$ during sintering. First,  Bourgeois \textit{et al.} proposed that a eutectic between metallic Cr and $\text{Cr}_2\text{O}_3$ lead to the formation of a liquid phase with 1:1 Cr to O stoichiometry that facilitates liquid phase sintering, responsible for grain growth \cite{BOURGEOIS2001313}. In this mechanism, Cr enriches at the GB and locally melts under sintering conditions, accelerating diffusion across the GB necessary for grain growth.
The basis for this mechanism is the fact that the most considerable grain growth occurred at the temperature and oxygen chemical potential closest to the domain of CrO stability and not in the regions of Cr and $\text{Cr}_2\text{O}_3$. Moreover, Bourgeois \textit{et al.} provide an optical micrograph (Figure \ref{fig:microscopy} B) of a secondary phase at a GB consisting of separated $\text{Cr}_2\text{O}_3$ and metallic Cr with a rounded morphology that points to this secondary phase forming from a quenched liquid. This disassociated secondary phase was also directly observed by Riglet-Martial \textit{et al.} under similar thermodynamic conditions predicted by the phase diagram \cite{RIGLETMARTIAL201463}. A Cr rich precipitate at GBs with mixed charge state has been widely reported \cite{devillaire2023characterisation,MURPHY2023}. Moreover, Yang \textit{et al.} propose that a stepwise sintering mechanism maximizes the concentration of Cr ions at the GB responsible for grain growth and attributes the abnormal grain growth to differences in initial concentration \cite{yang2012effect}. While enhanced grain growth is not precisely tied to the region of bulk CrO stability in Figure \ref{fig:phasediagram} A), the differential thermal analysis of Cr doped $\text{UO}_2$ has identified a peak associated with the recrystallization of liquid CrO at a lower oxygen potential and temperature than bulk $\text{Cr}_2\text{O}_3$ \cite{peres2012high}. This supports the claim that the CrO phase is stabilized by the surrounding $\text{UO}_2$. A recent study by Vallely \textit{et al.} presents differential scanning calorimetry curves for Cr-doped $\ch{UO_2}$ that appear to be linear combinations of $\ch{UO_2}$ and some $\ch{CrUO_4}$ features \cite{VALLELY2025156105}. In conjunction with enhanced grain growth and densification starting around \SI{1250}{\celsius}, their findings are suggestive that the Cr-O phase responsible for liquid sintering could be $\ch{CrUO_4}$. Further study is required to explore the potential phases within the Cr-O-U system that lead to enhanced grain growth in Cr-doped $\ch{UO_2}$. The liquid CrO sintering hypothesis explains most of the literature data on enhanced grain growth, with the largest grain sizes being found after sintering near the region of CrO stability. However, a CrO liquid phase might not be the only mechanism for enhancing the grain growth rate of Cr doped $\text{UO}_2$ that might be working synergistically. 

The alternative mechanisms to CrO doping focus on Cr doping, promoting U self-diffusivity by increasing the concentration of point defects like U vacancies. Evidence that Cr doping promotes the diffusivity of species on the $\text{UO}_2$ lattice at reactor relevant temperatures is provided in section \ref{FGR}, where the work of Killeen and Arborelius \textit{et al.} showed a competition between larger grain sizes in Cr doped $\text{UO}_2$ and higher fission gas diffusivity proportional to the concentration of $\text{Cr}_2\text{O}_3$ added to the fuel \cite{KILLEEN1980177,arborelius2006advanced}. There, we introduced two classes of mechanisms that could be responsible for enhanced diffusivity: (i) point defect interactions and (ii) bulk redox equilibrium. These two classes of mechanisms are tied to the question of how Cr is incorporated into the fuel dissolved into the lattice or precipitated as a second phase at GBs, corresponding to the two classes of mechanisms, respectively. 

Let us consider the first case of point defect interaction mechanisms. Some Cr will be dissolved on the  Cr doped $\text{UO}_2$ lattice as point defects either substitutionally for the $\text{U}^{+4}$ cation or on an interstitial site \cite{cardinaels2012chromia}. Atomistic modeling evidence \cite{CARDINAELS2012289, guo2017atomic} and spectroscopic analysis \cite{SMITH2022,MURPHY2023} point to Cr replacing U as a substitution type defect. The charge state of Cr in the U site is debated, with some \textit{ab initio} studies predicting a +2 speciation \cite{SUN2020} in contrast to spectroscopy results that report a +3 charge state \cite{MURPHY2023}, but for the purposes of defect chemistry, as substitutions have similar effects. The net negative charge of the proposed substitution mechanisms promotes the concentration of positively charged defects like oxygen vacancies and $\text{U}^{+5}$ (the latter mostly ruled out by spectroscopy), and decreases the concentration of negatively charged defects like U vacancies. A lower U vacancy concentration would decrease the diffusivity of U. Given the low concentration of Cr on the lattice, this effect is likely to be small, but Cr incorporating substitutionally as a negatively charged defect cannot be responsible for an increase in diffusivity. 

The second mechanism assumed that positively charged Cr defects are incorporated as either an interstitial or substitution in the +6 oxidation state \cite{MATSUI1986212}. Simply, the oxidation of Cr from the +3 to +6 oxidation state is energetically unfavorable and hence unlikely as a mechanism \cite{COOPER2018403}. On the basis of \textit{ab initio} calculations of defect formation energies, Cooper \textit{et al.} proposed by assuming a lower oxidation state \textit{e.g.} $\text{Cr}^{+1}_i$, Cr and other transition metals could incorporate interstitially at high enough concentrations in $\text{UO}_2$ to promote the concentration of U vacancies  \cite{COOPER2018403}. These increased vacancy concentrations were in turn used in grand potential simulations to show enhanced grain growth in doped $\ch{UO_2}$ \cite{GREENQUIST2020152052,greenquist2023phase}. However, the concentrations calculated by Cooper \textit{et al.} were at an oxygen partial pressure of $10^{-20}$\SI{}{atm} $\text{O}_2$ which yields an oxygen chemical potential of \SI{-700}{kJ/mol} which is \SI{300}{kJ/mol} below most reported sintering environments as shown in Figure \ref{fig:phasediagram}. In such low oxygen potential, Cr solubility in the lattice is negligible \cite{RIGLETMARTIAL201463} and grain growth is not observed \cite{yang2012effect}. In such reducing sintering environments, the equilibrium U vacancy concentration is several orders of magnitude lower than at oxygen potentials commonly used \cite{NIELSON2024}. Moreover, the effect of Cr point defects on the bulk concentrations of other point defects at sintering conditions would require concentrations approaching unity, which are nonphysical \cite{COOPER2018403}. However, a recent computational study has shown that favorable segregation of Cr interstitial to GBs can enhance the diffusivity via an analogous mechanism \cite{ClevelandChargestate}. Moreover, the sintering of Cr doped $\ch{UO_2}$ is  reported under thermodynamic conditions where the equilibrium O/U ratio is above 2 as seen in Figure \ref{fig:O/M ratio versus Grain Size} and Figure \ref{fig:phasediagram} B). It should be noted that at O/U ratios approaching 2, no acceleration in grain growth is reported \cite{yang2024novel}. Furthermore, the O/U ratio calculated in \ref{fig:phasediagram} B) neglects the contribution of the Cr/$\ch{Cr_2O_3}$ equilibrium in controlling the oxygen potential. The O/U ratio for a typical Cr doped $\ch{UO_2}$ pellet is greater than 2.005. 

While the minute concentration of $\text{Cr}_2\text{O}_3$ added and low solubility on the $\text{UO}_2$ lattice means that point defect models cannot explain the enhanced diffusivity, bulk thermodynamic mechanisms perform better. In the mechanism proposed by Cooper \textit{et al.} \cite{COOPER2021152590}, the equilibrium between Cr and $\text{Cr}_2\text{O}_3$ controls the oxygen potential under reactor conditions, enhancing the concentration of defect clusters responsible for fission gas diffusivity. In this model, Cr-rich precipitates determine the oxygen potential in the fuel, which in turn tunes the concentration of defects. This mechanism can explain the enhanced diffusivity as the higher oxygen chemical potential promotes hyper-stoichiometric defects, including U vacancies, resulting in higher U diffusivity and thus faster sintering \cite{COOPER2018251}. This increase in diffusivity by a factor of about 3 is sufficient to explain the larger grain sizes obtained when sintering in dry conditions, with higher Cr concentrations promoting the oxygen chemical potential. Therefore, the literature illustrates that Cr doping promotes the mobility of $\text{UO}_2$ grains primarily through a liquid phase sintering mechanism at oxidative sintering conditions and perhaps secondarily through the equilibrium between Cr/$\text{Cr}_2\text{O}_3$ which promotes the bulk diffusivity of species by raising the oxygen potential. 

\section{Future Directions}


\subsection{Irradiation behavior at high burnup}
Adding $\ch{Cr_2O_3}$ to $\ch{UO_2}$ can effectively reduce FGR during reactor operation. This ability is significant because it may allow fuel to operate in higher burnup regimes, enhancing both economic and resource efficiency. At high burnup, the fuel restructures with microstructure depending on the radial distance from the fuel rod core, with the outermost region pulverizing \cite{RONDINELLA201024,NOIROT2008318}. There has been much recent interest in understanding the micro-structural changes in fuels after high burnups in doped fuel with studies looking at transient behavior \cite{ zhao2024transient}, restructuring \cite{noirot2022restructuring}, mechanical properties \cite{zhaoeffects}, and oxidation resistance \cite{milena2025oxidation,milena2025raman} of doped and undoped fuels at high burnup. A mechanistic understanding of a dopants effect on the high burnup structure is limited by the current understanding of the causes of restructuring at high burnup; the role that enhanced diffusion and larger grain size plays can be studied. Planned irradiation experiments promise to remedy gaps in the understanding of FGR behavior left by the publicly available literature and better parameterize these models \cite{gorton2024modeling}.

\subsection{Modeling and experimental validation of the variable charge states}
The properties responsible for the enhanced grain growth of Cr-doped $\ch{UO_2}$ stem from the redox of the Cr ion between its +3, +2, and +0 oxidation states. While this system can be investigated via first principles simulations such as DFT \cite{COOPER2018403,SUN2020}, the computational cost of these calculations limits their size, hindering these methods from investigating behaviors on the length scales of GBs. Previous studies have worked around these computational limitations by using empirical interatomic potentials of a given functional form fit to a particular oxide structure with a fixed charge state \cite{MIDDLEBURGH2012258,COOPER2013236}. Moreover, the current state of the art relies on these empirical interatomic potentials to calculate the vibrational entropy changes of defective lattices \cite{COOPER2018251,NIELSON2024}. This approach might be suitable for undoped $\ch{UO}_2$, it presents a significant problem because the calculation of the free energies of charged defects involving Cr and other transition metal ions that can assume multiple charge states relies on the extrapolation of an empirical potential with a fixed Cr +3 charge state \cite{COOPER2018403,OWEN2023154270}. While the fixed charge state potential is a valuable scientific tool for the investigation of the primary mode of Cr incorporation into the lattice and its impact on diffusivity and lattice parameter \cite{owen2022role} as well as GB segregation energies \cite{TERRICABRAS2025156114}, further advancements are needed to model the variable charge states with an approach more computationally efficient than first principles calculations. Machine learned interatomic potentials have shown promise in predicting multiple charge states in different systems and incorporating them into a long range electrostatic framework \cite{PANNAv2}, and it can be extended to doped $\ch{UO_2}$ fuel.

The variable charge state has been shown by Murphy \textit{et al.} with some success \cite{MURPHY2023}. However, there are still several unknown variables that should be accounted for while determining the charge state in the bulk and at the GBs. For instance, given that the local atomic environments of the GB are influenced by the GB type and sintering temperature. The task that lies ahead includes correlating the charge states with specific or dominant GBs that have been observed in $\ch{UO_2}$ fuel. One approach is to do a bulk to GB sweep using electron energy loss spectroscopy (EELS). Bawane \textit{et al.} showed the use of EELS to determine the oxidation of a nickel alloy after exposure to molten salts \cite{BAWANE2021113790}. XANES measurements can offer insights into the local electronic structure. However, obtaining site-specific information comparable to EELS requires harvesting samples specifically from the GB region. This process necessitates sample preparation guided by optical microscopy techniques.


\subsection{Direct observation of grain boundary potential}
One element of understanding the Cr rich precipitates that form at GB phases is understanding their impact on the electronic properties of $\ch{UO_2}$. While a previous study found that Cr doping lowers the electrical conductivity by an order of magnitude \cite{MATSUI1986212} the mechanism is uncertain. Recent computational work has theorized that the segregation of charged defects to the $\ch{UO_2}$ GB leads to the formation of net electrostatic charge that changes the defect chemistry around the GBs \cite{Wolf2025}. In our previous work, we suggested how Cr GB segregation would effect this space charge effect as a function of GB structure and dopant charge state \cite{ClevelandChargestate}. However, modeling efforts are limited, and the proposed GB charge must be investigated experimentally. While previous efforts using a method like traditional impedance spectroscopy \cite{KIM2024116706} and TEM have not been able to identify space charge in $\ch{UO_2}$ despite being able to identify it in other similar oxide {\cite{toyama2023real}, method like electron ptychography that can image the GB electric field with atomic precision is needed \cite{toyama2024direct}. Gilgenbach \textit{et al.} successfully imaged space charge layers in $\ch{CeO_2}$ using electron ptychography as shown in Figure \ref{fig:LeBeau} \cite{Gilgenbach2024} which agrees with simulations.

\begin{figure}
    \centering
    \includegraphics[width=1\linewidth]{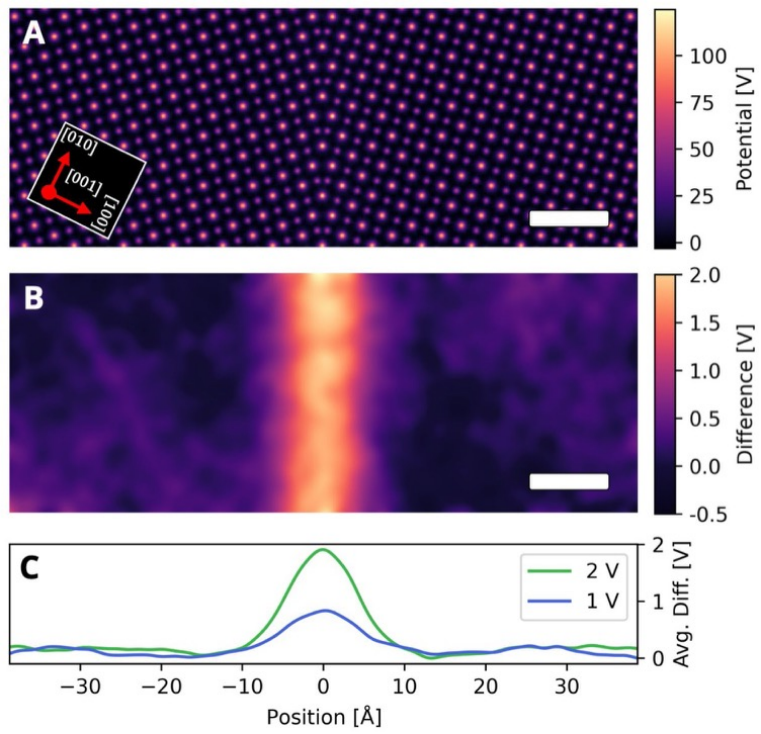}
    \caption{Simulated multislice ptychographic reconstruction of a $\Sigma5$ symmetric tilt grain boundary in $\ch{CeO_2}$. (A) Ptychographic reconstruction with a 2 V peak, 1 nm FWHM Gaussian applied space charge potential, displayed as a thickness-averaged potential. (B) Difference between reconstruction with and without the 2 V space charge potential (a 1 Å Gaussian blur has been applied). (C) A linescan of (B) for two different values of applied potential. Scale bars 1 nm. Taken from GilgenBach \textit{et al} \cite{Gilgenbach2024}}
    \label{fig:LeBeau}
\end{figure}

An alternative approach is to look at the mechanical creep properties of doped $\ch{UO_2}$ and infer from those defect concentrations to compare with theory \cite{galvin2022development}. Integrating atomic resolution characterization with theoretical predictions offers to advance the frontier of our understanding of Cr doped $\ch{UO_2}$ microstructure and dopant chemistry. 

\section{Conclusions}

$\text{UO}_2$ doped with 500-2000 ppm wt\% $\text{Cr}_2\text{O}_3$ is considered a promising near term fuel concept for large grained accident tolerant nuclear fuels. While the FGR performance of Cr-doped $\text{UO}_2$ is governed by a competition between enhanced lattice diffusivity and the resulting larger grain size of Cr-doped $\text{UO}_2$. Sintering under conditions to produce a Cr/$\text{Cr}_2\text{O}_3$ eutectic can achieve larger grain sizes that outweigh the effect of fast fission gas diffusion due to the change in the oxygen potential of the fuel. The narrow range of thermodynamic stability of the Cr-rich GB phase enhances diffusion under sintering conditions, but not under reactor conditions where oxygen potential effects dominate. Therefore, Cr-doped $\text{UO}_2$ can be considered analogous to $\text{UO}_2$, but with larger grain sizes and larger diffusivity. Efforts to model this in fuel performance codes have shown significant promise in facilitating the certification of new fuel designs. While more work needs to be done to fully understand the behavior of irradiated doped fuels under long term environmental storage, the initial results point to chemical stability similar to or improved upon that of  $\text{UO}_2$ fuels. The stage is set for Cr-doped $\text{UO}_2$ fuels to reach technological maturity, where their benefits can be leveraged to produce safer and more efficient nuclear power.

\section{Acknowledgments }
This work was funded by the Faculty Startup Fund and Research Support Committee grant from the Massachusetts Institute of Technology. 

\bibliographystyle{unsrt}
\bibliography{sample}

\begin{thebibliography}{100}

\bibitem{Belle_UO2_1961}
J.~Belle.
\newblock {\em Uranium dioxide: properties and nuclear applications.}
\newblock Naval Reactors, Division of Reactor Development, U.S. Atomic Energy Commission, Washington, 1961.

\bibitem{sadik_evaluation_1959}
Kakac Sadik.
\newblock Evaluation of {Uranium} {Dioxide} as a {Fuel} {Element} in {Nuclear} {Reactors}.
\newblock Master's thesis, MIT, Barker Engineering Library, 1959.

\bibitem{hewlett_new_1962}
Richard Hewlett and Oscar Anderson.
\newblock {\em The {New} {World}, 1939/1946, {A} {History} of {The} {United} States {Atomic} {Energy} {Commission}, {Volume} {I}}, volume Volume 1.
\newblock The Pennsylvania State University Press, 1962.

\bibitem{LYONS1972167}
M.F. Lyons, R.F. Boyle, J.H. Davies, V.E. Hazel, and T.C. Rowland.
\newblock $\ch{UO_2}$ properties affecting performance.
\newblock {\em Nuclear Engineering and Design}, 21(2):167--199, 1972.

\bibitem{mceachern_review_1998}
R.J McEachern and P~Taylor.
\newblock A review of the oxidation of uranium dioxide at temperatures below 400°{C}.
\newblock {\em Journal of Nuclear Materials}, 254(2-3):87--121, 1998.

\bibitem{FINK20001}
J.K. Fink.
\newblock Thermophysical properties of uranium dioxide.
\newblock {\em Journal of Nuclear Materials}, 279(1):1--18, 2000.

\bibitem{kolstad_fuel_1992}
Eric Kolstad and Carlo Vitanza.
\newblock Fuel rod and core materials investigations related to {LWR} extended burnup operation.
\newblock {\em Journal of Nuclear Materials}, 188, June 1992.

\bibitem{COOPER2021152590}
Michael~WD Cooper, Giovanni Pastore, Yifeng Che, Christopher Matthews, Axel Forslund, Christopher~R Stanek, Koroush Shirvan, Terje Tverberg, Kyle~A Gamble, Brian Mays, and David~A Andersson.
\newblock Fission gas diffusion and release for $\ch{Cr_2O_3}$-doped $\ch{UO_2}$: From the atomic to the engineering scale.
\newblock {\em Journal of Nuclear Materials}, 545:152590, 2021.

\bibitem{rest_fission_2019}
J~Rest, M.W.D Cooper, J~Spino, J.A. Turnbull, P~Van~Uffelen, and C.~T. Walker.
\newblock Fission gas release from {$\ch{UO_2}$} nuclear fuel: {A} review.
\newblock {\em Journal of Nuclear Materials}, 513:310--345, 2019.

\bibitem{BOOTH1957}
A.H Booth.
\newblock A method of calculating fission gas diffusion from $\ch{UO_2}$ fuel and its application to the x-2-f loop test.
\newblock {\em AECL}, 1957.

\bibitem{ASTM_C753-16}
ASTM C753-16.
\newblock Standard speciﬁcation for nuclear-grade, sinterable uranium dioxide powder1, 2021.

\bibitem{HARADA1997217}
Y.~Harada.
\newblock $\ch{UO_2}$ sintering in controlled oxygen atmospheres of three-stage process.
\newblock {\em Journal of Nuclear Materials}, 245(2):217--223, 1997.

\bibitem{BURKE1952220}
J.E. Burke and D.~Turnbull.
\newblock Recrystallization and grain growth.
\newblock {\em Progress in Metal Physics}, 3:220--292, 1952.

\bibitem{doi1964significance}
H~Doi and T~Ito.
\newblock Significance of physical state of starting precipitate in growth of uranium dioxide particles.
\newblock {\em Journal of Nuclear Materials}, 11(1):94--106, 1964.

\bibitem{https://doi.org/10.1111/j.1151-2916.1963.tb11692.x}
N.~FUHRMAN, L.~D. HOWER~JR., and R.~B. HOLDEN.
\newblock Low-temperature sintering of uranium dioxide.
\newblock {\em Journal of the American Ceramic Society}, 46(3):114--121, 1963.

\bibitem{kashibe1998effect}
S~Kashibe and K~Une.
\newblock Effect of additives ($\ch{Cr_2O_3}$, $\ch{Al_2O_3}$,$\ch{SiO_2}$,$\ch{MgO}$) on diffusional release of 133$\ch{Xe}$ from $\ch{UO_2}$ fuels.
\newblock {\em Journal of nuclear materials}, 254(2-3):234--242, 1998.

\bibitem{Winter1965}
K.~Winter.
\newblock Effect of chromium(iii)-oxide on the sintering properties of uranium dioxide.
\newblock {\em Kernenergie (East Germany)}, 8, 1965-07-01.

\bibitem{etde_20628373}
D~Ohai.
\newblock Large grain size $\ch{UO_2}$ sintered pellets obtaining used for burn up extension, Jul 2003.

\bibitem{TURNBULL197462}
J.A. Turnbull.
\newblock The effect of grain size on the swelling and gas release properties of $\ch{UO_2}$ during irradiation.
\newblock {\em Journal of Nuclear Materials}, 50(1):62--68, 1974.

\bibitem{zinkle_accident_2014}
S.J. Zinkle, K.A. Terrani, J.C. Gehin, L.J. Ott, and L.L. Snead.
\newblock Accident tolerant fuels for {LWRs}: {A} perspective.
\newblock {\em Journal of Nuclear Materials}, 448(1-3):374--379, May 2014.

\bibitem{key}
Luther~H. Hallman.
\newblock Westinghouse advanced doped pellet technology (adopttm) fue.
\newblock \url{https://www.nrc.gov/docs/ML2013/ML20132A014.pdf}, 2020.
\newblock [Accessed 16-09-2025].

\bibitem{ADOPT_NRC}
Final safety evaluation by the office of nuclear reactor regulation for the westinghouse electric company topical report wcap-18482-p/wcap-18482-np, revision 0, “westinghouse advanced doped pellet technology (adopt™) fuel” docket no. 99902038 epid l-2020-top-0025, 2020.

\bibitem{Lewis1957}
\url{For a comprehensive description of the new fuel form see W.B. Lewis, The Significance of Developing a High Performance Uranium Oxide Fuel, AECLDM-44, Publication No. 478, 1957}.
\newblock [Accessed 15-09-2025].

\bibitem{TERRICABRAS2025156114}
Adrien~J. Terricabras, Conor~O.T. Galvin, Maria Kosmidou, Miguel Pena, Arjen {van Veelen}, William~D. Neilson, Shen~J. Dillon, Michael~W.D. Cooper, David~A. Andersson, Sarah~C. Finkeldei, and Joshua~T. White.
\newblock Chromium doping effects on uo2 grain boundary chemistry: A combined experimental and modeling approach.
\newblock {\em Journal of Nuclear Materials}, page 156114, 2025.

\bibitem{MIDDLEBURGH2023154250}
Simon~C. Middleburgh, Simon Dumbill, Adam Qaisar, Ian Vatter, Megan Owen, Sarah Vallely, Dave Goddard, David Eaves, Mattias Puide, Magnus Limbäck, and William~E. Lee.
\newblock Enrichment of chromium at grain boundaries in chromia doped $\ch{UO_2}$.
\newblock {\em Journal of Nuclear Materials}, 575:154250, 2023.

\bibitem{Murphy2024}
Gabriel Murphy, Elena Bazarkina, Andre Rossberg, and et~al.
\newblock Elucidating the roles of redox and structure in the grain growth of mn-doped $\ch{UO_2}$.
\newblock {\em ReserachSquare}, 2024.

\bibitem{Powers1960}
\url{Powers, Richard M. "The effect of solid solution additions on the thermal conductivity of UO2." Vol. 317. Technical Information Service, 1960.}
\newblock [Accessed 15-09-2025].

\bibitem{ANG1960176}
C.Y. Ang and K.W. Burkhammer.
\newblock Sintering of high density uranium dioxide bodies.
\newblock {\em Journal of Nuclear Materials}, 2(2):176--180, 1960.

\bibitem{ainscough_rigby_osborn_1974}
J.B. Ainscough, F.~Rigby, and S.C. Osborn.
\newblock The effect of titania on grain growth and densification of sintered uo2.
\newblock {\em J. Nucl. Mater.}, 52(2):191–203, Oct 1974.

\bibitem{AMATO1966252}
I.~Amato, R.L. Colombo, and A.Petruccioli Balzari.
\newblock Grain growth in pure and titania-doped uranium dioxide.
\newblock {\em Journal of Nuclear Materials}, 18(3):252--260, 1966.

\bibitem{German01042013}
R~M German.
\newblock History of sintering: empirical phase.
\newblock {\em Powder Metallurgy}, 56(2):117--123, 2013.

\bibitem{hannay2012treatise}
N.~Hannay.
\newblock {\em Treatise on Solid State Chemistry: Volume 4 Reactivity of Solids}.
\newblock Chemistry and Materials Science. Springer US, 2012.

\bibitem{PAIK201334}
Shrishma Paik, S.~Biswas, S.~Bhattacharya, and S.B. Roy.
\newblock Effect of ammonium nitrate on precipitation of ammonium di-uranate (adu) and its characteristics.
\newblock {\em Journal of Nuclear Materials}, 440(1):34--38, 2013.

\bibitem{googleUS4430276AMethod}
{U}{S}4430276{A} - {M}ethod of making stable {U}{O}2 fuel pellets - {G}oogle {P}atents --- patents.google.com.
\newblock \url{https://patents.google.com/patent/US4430276A/en}.
\newblock [Accessed 16-09-2025].

\bibitem{googleGB1285190AImprovements}
{G}{B}1285190{A} - {I}mprovements in ceramic fissile materials - {G}oogle {P}atents --- patents.google.com.
\newblock \url{https://patents.google.com/patent/GB1285190A/en}.
\newblock [Accessed 16-09-2025].

\bibitem{toraji1968method}
Nishijima Toraji, Fukushima Eiji, Ishihata Akira, Wada Takashi, Kawada Toshiyuki, and Kitagawa Haruya.
\newblock Method of preparing nuclear fuel pellets, February~27 1968.
\newblock US Patent 3,371,133.

\bibitem{matzke_atomic_1966}
Hj. Matzke and Ontario~(Canada) Atomic Energy~of Canada~Limited, Chalk~River.
\newblock On the effect of tio<sub>2</sub> additions on defect structure, sintering and gas release of uo<sub>2</sub>.
\newblock May 1966.

\bibitem{UNE198793}
Katsumi Une, Isami Tanabe, and Masaomi Oguma.
\newblock Effects of additives and the oxygen potential on the fission gas diffusion in uo2 fuel.
\newblock {\em Journal of Nuclear Materials}, 150(1):93--99, 1987.

\bibitem{googleFR2706066B1Nuclear}
{F}{R}2706066{B}1 - {N}uclear fuel with improved fission product retention properties. - {G}oogle {P}atents --- patents.google.com.
\newblock \url{https://patents.google.com/patent/FR2706066B1/en?oq=FR2706066B1+(1988)}.
\newblock [Accessed 16-09-2025].

\bibitem{KLEYKAMP1997103}
Heiko Kleykamp.
\newblock Phase equilibria in the $\ch{UO_2}$-austenitic steel system up to 3000°c.
\newblock {\em Journal of Nuclear Materials}, 247:103--107, 1997.
\newblock Thermodynamics of Nuclear Materials.

\bibitem{leenaers2003solubility}
A~Leenaers, L~De~Tollenaere, Ch~Delafoy, and S~Van~den Berghe.
\newblock On the solubility of chromium sesquioxide in uranium dioxide fuel.
\newblock {\em Journal of nuclear materials}, 317(1):62--68, 2003.

\bibitem{cardinaels2012chromia}
Thomas Cardinaels, Kevin Govers, Benedict Vos, Sven Van~den Berghe, Marc Verwerft, L~De~Tollenaere, G~Maier, and C~Delafoy.
\newblock Chromia doped $\ch{UO_2}$ fuel: Investigation of the lattice parameter.
\newblock {\em Journal of nuclear materials}, 424(1-3):252--260, 2012.

\bibitem{BOURGEOIS2001313}
L.~Bourgeois, Ph. Dehaudt, C.~Lemaignan, and A.~Hammou.
\newblock Factors governing microstructure development of $\ch{Cr_2O_3}$-doped $\ch{UO_2}$ during sintering.
\newblock {\em Journal of Nuclear Materials}, 297(3):313--326, 2001.

\bibitem{RIGLETMARTIAL201463}
Ch. Riglet-Martial, Ph. Martin, D.~Testemale, C.~Sabathier-Devals, G.~Carlot, P.~Matheron, X.~Iltis, U.~Pasquet, C.~Valot, C.~Delafoy, and R.~Largenton.
\newblock Thermodynamics of chromium in $\ch{UO_2}$ fuel: A solubility model.
\newblock {\em Journal of Nuclear Materials}, 447(1):63--72, 2014.

\bibitem{milena2021raman}
A~Milena-P{\'e}rez, LJ~Bonales, Nieves Rodr{\'\i}guez-Villagra, S~Fern{\'a}ndez, Valent{\'\i}n~G Baonza, and J~Cobos.
\newblock Raman spectroscopy coupled to principal component analysis for studying $\ch{UO_2}$ nuclear fuels with different grain sizes due to the chromia addition.
\newblock {\em Journal of Nuclear Materials}, 543:152581, 2021.

\bibitem{MURPHY2023}
Gabriel~L. Murphy, Robert Gericke, Sara Gilson, Elena~F. Bazarkina, André Rossberg, Peter Kaden, Robert Thümmler, Martina Klinkenberg, Maximilian Henkes, Philip Kegler, Volodymyr Svitlyk, Julien Marquardt, Theresa Lender, Christoph Hennig, Kristina~O. Kvashnina, and Nina Huittinen.
\newblock Deconvoluting $\ch{Cr}$ states in $\ch{Cr}$-doped $\ch{UO_2}$ nuclear fuels via bulk and single crystal spectroscopic studies.
\newblock {\em Nature Communications}, 14:2041--1723, 2023.

\bibitem{MIDDLEBURGH2012258}
S.C. Middleburgh, D.C. Parfitt, R.W. Grimes, B.~Dorado, M.~Bertolus, P.R. Blair, L.~Hallstadius, and K.~Backman.
\newblock Solution of trivalent cations into uranium dioxide.
\newblock {\em Journal of Nuclear Materials}, 420(1):258--261, 2012.

\bibitem{Gascoin2025}
Mathieu Gascoin, Mariya Romanova, Ibrahim~Cheik Njifon, and Michel Freyss.
\newblock Dft+u investigation of local configurations and oxidation states of cr in cr-doped $\ch{UO_2}$.
\newblock {\em Communications Chemistry}, 8:257, 2025.

\bibitem{COOPER2013236}
M.W.D. Cooper, D.J. Gregg, Y.~Zhang, G.J. Thorogood, G.R. Lumpkin, R.W. Grimes, and S.C. Middleburgh.
\newblock Formation of $\ch{(Cr,Al)UO_4}$ from doped $\ch{UO_2}$ and its influence on partition of soluble fission products.
\newblock {\em Journal of Nuclear Materials}, 443(1):236--241, 2013.

\bibitem{COOPER2018403}
M.W.D. Cooper, C.R. Stanek, and D.A. Andersson.
\newblock The role of dopant charge state on defect chemistry and grain growth of doped $\ch{UO_2}$.
\newblock {\em Acta Materialia}, 150:403--413, 2018.

\bibitem{ClevelandChargestate}
Mack Cleveland and Ericmoore Jossou.
\newblock Effects of charge state on transition metal segregation at grain boundaries in uranium dioxide.
\newblock Manuscript in Preparation.

\bibitem{kegler2020chromium}
P~Kegler, M~Klinkenberg, A~Bukaemskiy, F~Brandt, G~Deissmann, and D~Bosbach.
\newblock Chromium doped $\ch{UO_2}$-based model systems: Synthesis and characterization of model materials for the study of the matrix corrosion of spent modern nuclear fuels.
\newblock In {\em 2 nd Annual Meeting Proceedings}, page~41, 2020.

\bibitem{TERRICABRAS2024155022}
Adrien~J. Terricabras, Sean~M. Drewry, Keri Campbell, Elizabeth~J. Judge, Darrin~D. Byler, Emily~S. Teti, Arjen {van Veelen}, Scarlett {Widgeon Paisner}, and Joshua~T. White.
\newblock Performance and properties evolution of near-term accident tolerant fuel: $\ch{Cr}$-doped $\ch{UO_2}$.
\newblock {\em Journal of Nuclear Materials}, 594:155022, 2024.

\bibitem{SMITH2022}
Hannah Smith, Luke~T. Townsend, Ritesh Mohun, Théo Cordara, J.~Frederick~W. Mosselmans, Kristina Kvashnina, and Claire~L. Corkhill.
\newblock Cr2+ solid solution in $\ch{UO_2}$ evidenced by advanced spectroscopy.
\newblock {\em Communications Chemistry}, 5:2399--3669, 2022.
\newblock PMID: 37713315.

\bibitem{CARDINAELS2012289}
T.~Cardinaels, J.~Hertog, B.~Vos, L.~{de Tollenaere}, C.~Delafoy, and M.~Verwerft.
\newblock Dopant solubility and lattice contraction in gadolinia and gadolinia–chromia doped $\ch{UO_2}$ fuels.
\newblock {\em Journal of Nuclear Materials}, 424(1):289--300, 2012.

\bibitem{silva2021evaluation}
Chinthaka~M Silva, Rodney~D Hunt, and Kiel~S Holliday.
\newblock An evaluation of tri-valent oxide ($\ch{Cr_2O_3}$) as a grain enlarging dopant for $\ch{UO_2}$ nuclear fuels fabricated under reducing environment.
\newblock {\em Journal of Nuclear Materials}, 553:153053, 2021.

\bibitem{SILVA2021153003}
Chinthaka~M. Silva, Rodney~D. Hunt, and Andrew~T. Nelson.
\newblock Microstructural and crystallographic effects of sol-gel synthesized ti-doped $\ch{UO_2}$ sintered under reducing conditions.
\newblock {\em Journal of Nuclear Materials}, 552:153003, 2021.

\bibitem{SUN2020}
M.~Sun, J.~Stackhouse, and P.M. Kowalski.
\newblock The +2 oxidation state of $\ch{Cr}$ incorporated into the crystal lattice of $\ch{UO_2}$.
\newblock {\em Communications Materials}, 1:2662--4443, 2020.
\newblock PMID: 37713315.

\bibitem{mieszczynski2014microbeam}
C~Mieszczynski, G~Kuri, J~Bertsch, M~Martin, CN~Borca, Ch~Delafoy, and E~Simoni.
\newblock Microbeam x-ray absorption spectroscopy study of chromium in large-grain uranium dioxide fuel.
\newblock {\em Journal of Physics: Condensed Matter}, 26(35):355009, 2014.

\bibitem{guo2017atomic}
Zhexi Guo, Raoul Ngayam-Happy, Matthias Krack, and Andreas Pautz.
\newblock Atomic-scale effects of chromium-doping on defect behaviour in uranium dioxide fuel.
\newblock {\em Journal of Nuclear Materials}, 488:160--172, 2017.

\bibitem{PANNAv2}
Franco Pellegrini, Ruggero Lot, Yusuf Shaidu, and Emine Küçükbenli.
\newblock Panna 2.0: Efficient neural network interatomic potentials and new architectures.
\newblock {\em The Journal of Chemical Physics}, 159(8):084117, 08 2023.

\bibitem{moore2024high}
Guy~C Moore, Matthew~K Horton, Edward Linscott, Alexander~M Ganose, Martin Siron, David~D O'Regan, and Kristin~A Persson.
\newblock High-throughput determination of hubbard u and hund j values for transition metal oxides via the linear response formalism.
\newblock {\em Physical Review Materials}, 8(1):014409, 2024.

\bibitem{yang2012effect}
Jae~Ho Yang, Keon~Sik Kim, Ik~Hui Nam, Jang~Soo Oh, Dong-Joo Kim, Young~Woo Rhee, and Jong~Hun Kim.
\newblock Effect of step wise variation of oxygen potential during the isothermal sintering on the grain growth behavior in $\ch{Cr_2O_3}$ doped $\ch{UO_2}$ pellets.
\newblock {\em Journal of nuclear materials}, 429(1-3):25--33, 2012.

\bibitem{zhong2021preparation}
Yi~Zhong, Rui Gao, Bingqing Li, Zhenliang Yang, Qiqi Huang, Zhiyi Wang, Limei Duan, Xuxu Liu, Mingfu Chu, Pengcheng Zhang, et~al.
\newblock Preparation and characterization of large grain $\ch{UO_2}$ for accident tolerant fuel.
\newblock {\em Frontiers in Materials}, 8:651074, 2021.

\bibitem{watanabe2023oxygen}
Masashi Watanabe and Masato Kato.
\newblock Oxygen potential, oxygen diffusion, and defect equilibria in $\ch{UO_2}$$\pm$x.
\newblock {\em Frontiers in Nuclear Engineering}, 1:1082324, 2023.

\bibitem{toker1991equilibrium}
NY~Toker, LS~Darken, and Arnulf Muan.
\newblock Equilibrium phase relations and thermodynamics of the $\ch{Cr-O}$ system in the temperature range of 1500° c to 1825° c.
\newblock {\em Metallurgical Transactions B}, 22:225--232, 1991.

\bibitem{peres2012high}
V{\'e}ronique Peres, Lo{\"\i}c Favergeon, Marie Andrieu, Jean-Claude Palussi{\`e}re, Julien Balland, Christine Delafoy, and Mich{\`e}le Pijolat.
\newblock High temperature chromium volatilization from $\ch{Cr_2O_3}$ powder and $\ch{Cr_2O_3}$-doped $\ch{UO_2}$ pellets in reducing atmospheres.
\newblock {\em Journal of Nuclear Materials}, 423(1-3):93--101, 2012.

\bibitem{devillaire2023characterisation}
Aubin Devillaire, Lionel Desgranges, Nicolas Tarisien, Xavi{\`e}re Iltis, Chantal Riglet-Martial, Ingrid Roure, Philippe Bienvenu, and Aurelien Canizares.
\newblock Characterisation of 3000 ppm $\ch{Cr_2O_3}$ doped $\ch{UO_2}$ and its precipitates.
\newblock {\em Journal of Raman Spectroscopy}, 54(5):562--568, 2023.

\bibitem{KURI2014158}
G.~Kuri, C.~Mieszczynski, M.~Martin, J.~Bertsch, C.N. Borca, and Ch. Delafoy.
\newblock Local atomic structure of chromium bearing precipitates in chromia doped uranium dioxide investigated by combined micro-beam x-ray diffraction and absorption spectroscopy.
\newblock {\em Journal of Nuclear Materials}, 449(1):158--167, 2014.

\bibitem{KILLEEN1980177}
J.C. Killeen.
\newblock Fission gas release and swelling in $\ch{UO_2}$ doped with $\ch{Cr_2O_3}$.
\newblock {\em Journal of Nuclear Materials}, 88(2):177--184, 1980.

\bibitem{KILLEEN197539}
J.C. Killeen.
\newblock The effect of additives on the irradiation behaviour of $\ch{UO_2}$.
\newblock {\em Journal of Nuclear Materials}, 58(1):39--46, 1975.

\bibitem{arborelius2006advanced}
Jakob Arborelius, Karin Backman, Lars Hallstadius, Magnus Limb{\"a}ck, Jimmy Nilsson, Bj{\"o}rn Rebensdorff, Gang Zhou, Koji Kitano, Reidar L{\"o}fstr{\"o}m, and Gunnar R{\"o}nnberg.
\newblock Advanced doped $\ch{UO_2}$ pellets in lwr applications.
\newblock {\em Journal of Nuclear Science and Technology}, 43(9):967--976, 2006.

\bibitem{OWEN2023154270}
Megan~W. Owen, Michael~W.D. Cooper, Michael~J.D. Rushton, Antoine Claisse, William~E. Lee, and Simon~C. Middleburgh.
\newblock Diffusion in undoped and $\ch{Cr}$-doped amorphous $\ch{UO_2}$.
\newblock {\em Journal of Nuclear Materials}, 576:154270, 2023.

\bibitem{CHE2018271}
Yifeng Che, Giovanni Pastore, Jason Hales, and Koroush Shirvan.
\newblock Modeling of $\ch{Cr_2O_3}$-doped $\ch{UO_2}$ as a near-term accident tolerant fuel for lwrs using the bison code.
\newblock {\em Nuclear Engineering and Design}, 337:271--278, 2018.

\bibitem{HALES2013531}
J.D. Hales, R.L. Williamson, S.R. Novascone, D.M. Perez, B.W. Spencer, and G.~Pastore.
\newblock Multidimensional multiphysics simulation of triso particle fuel.
\newblock {\em Journal of Nuclear Materials}, 443(1):531--543, 2013.

\bibitem{MATTHEWS2020152326}
Christopher Matthews, Romain Perriot, M.W.D Cooper, Christopher~R. Stanek, and David~A. Andersson.
\newblock Cluster dynamics simulation of xenon diffusion during irradiation in $\ch{UO_2}$.
\newblock {\em Journal of Nuclear Materials}, 540:152326, 2020.

\bibitem{gamble2019atf}
Kyle~A Gamble, Giovanni Pastore, David Andersson, and Michael~WD Cooper.
\newblock Atf material model development and validation for priority fuel concepts.
\newblock Technical report, Idaho National Lab.(INL), Idaho Falls, ID (United States), 2019.

\bibitem{tverberg2014update}
T~Tverberg.
\newblock Update on the in-pile results from the fission gas release mechanisms study in ifa-716.
\newblock {\em HWR-1090, Organisation for Economic Co-operation and Development, Halden Reactor Project}, 2014.

\bibitem{bremont2011ifa}
O~Br{\'e}mont.
\newblock Ifa-716.1 fission gas release mechanisms.
\newblock {\em HWR-1008, Organisation for Economic Cooperation and Development, Halden Reactor Project}, 2011.

\bibitem{che2021application}
Yifeng Che, Xu~Wu, Giovanni Pastore, Wei Li, and Koroush Shirvan.
\newblock Application of kriging and variational bayesian monte carlo method for improved prediction of doped $\ch{UO_2}$ fission gas release.
\newblock {\em Annals of Nuclear Energy}, 153:108046, 2021.

\bibitem{cheniour2023sensitivity}
Amani Cheniour, Ryan~T Sweet, Andrew~T Nelson, Brandon~A Wilson, and Ashley~E Shields.
\newblock Sensitivity of uo2 fuel performance to microstructural evolutions driven by dilute additives.
\newblock {\em Nuclear Engineering and Design}, 410:112383, 2023.

\bibitem{gorton2024modeling}
Jacob~P Gorton, Annabelle~G Le~Coq, Zane~G Wallen, Christian~M Petrie, Joshua~T White, John~T Dunwoody, Shane Mann, Nathan~A Capps, and Andrew~T Nelson.
\newblock Modeling and design of a separate effects irradiation test targeting fission gas release from $\ch{Cr}$-doped $\ch{UO_2}$.
\newblock {\em Nuclear Engineering and Design}, 429:113571, 2024.

\bibitem{nicodemo2024chromia}
Giovanni Nicodemo, Giovanni Zullo, Fabiola Cappia, Paul Van~Uffelen, Alejandra De~Lara, Lelio Luzzi, and Davide Pizzocri.
\newblock Chromia-doped $\ch{UO_2}$ fuel: An engineering model for chromium solubility and fission gas diffusivity.
\newblock {\em Journal of Nuclear Materials}, 601:155301, 2024.

\bibitem{backman2010westinghouse}
Karin Backman, Lars Hallstadius, and Gunnar R{\"o}nnberg.
\newblock Westinghouse advanced doped pellet-characteristics and irradiation behaviour.
\newblock {\em IAEA - Technical Meeting on Advanced Fuel Pellet Materials and Fuel Rod Designs for Water Cooled Reactors}, 2010.

\bibitem{fraczkiewicz2010study}
M~Fraczkiewicz.
\newblock Study of physical modifications induced by chromium doping of uranium dioxide.
\newblock Technical report, Grenoble Univ.(France), 2010.

\bibitem{mieszczynski2014irradiation}
C~Mieszczynski, G~Kuri, C~Degueldre, M~Martin, J~Bertsch, CN~Borca, D~Grolimund, Ch~Delafoy, and E~Simoni.
\newblock Irradiation effects and micro-structural changes in large grain uranium dioxide fuel investigated by micro-beam x-ray diffraction.
\newblock {\em Journal of nuclear materials}, 444(1-3):274--282, 2014.

\bibitem{massih2014effects}
Ali Massih.
\newblock {\em Effects of additives on uranium dioxide fuel behavior}.
\newblock Str{\aa}ls{\"a}kerhetsmyndigheten (SSM), 2014.

\bibitem{fink_thermophysical_2000}
J~Fink.
\newblock Thermophysical properties of uranium dioxide.
\newblock {\em Journal of Nuclear Materials}, 279(1):1--18, March 2000.

\bibitem{delafoy2006areva}
C~Delafoy and P~Dewes.
\newblock Areva np new $\ch{UO_2}$ fuel development and qualification for lwrs applications.
\newblock In {\em 2006 International LWR Fuel Performance Meeting (Top Fuel 2006), Salamanca, Spain}, 2006.

\bibitem{delafoy2018benefits}
C~Delafoy, J~Bischoff, J~Larocque, P~Attal, L~Gerken, and K~Nimishakavi.
\newblock Benefits of framatome’s e-atf evolutionary solution: Cr-coated cladding with $\ch{Cr_2O_3}$-doped fuel.
\newblock {\em Proceedings of the TopFuel}, pages 1--11, 2018.

\bibitem{cui2023thermodynamic}
Jinjiang Cui.
\newblock {\em Thermodynamic analysis of chemical interactions of the Zr-Cr-O and U-Cr-O systems at high temperatures for the evaluation of ATF (Accident Tolerant Fuel) under accidental conditions}.
\newblock PhD thesis, Universit{\'e} de Lille, 2023.

\bibitem{konarski2019thermo}
Piotr Konarski.
\newblock {\em Thermo-chemical-mechanical modeling of nuclear fuel behavior: Impact of oxygen transport in the fuel on Pellet Cladding Interaction}.
\newblock PhD thesis, Universit{\'e} de Lyon, 2019.

\bibitem{MILENAPEREZ2023154502}
A.~Milena-Pérez, L.J. Bonales, N.~Rodríguez-Villagra, M.B. Gómez-Mancebo, and H.~Galán.
\newblock Oxidation of accident tolerant fuels models based on $\ch{Cr}$-doped $\ch{UO_2}$ for the safety of nuclear storage facilities.
\newblock {\em Journal of Nuclear Materials}, 582:154502, 2023.

\bibitem{cachoir2021evolution}
Christelle Cachoir, Thierry Mennecart, and Karel Lemmens.
\newblock Evolution of the uranium concentration in dissolution experiments with cr-(pu) doped uo 2 in reducing conditions at sck cen.
\newblock {\em MRS Advances}, 6:84--89, 2021.

\bibitem{NILSSON2017123}
Kristina Nilsson, Olivia Roth, and Mats Jonsson.
\newblock Oxidative dissolution of adopt compared to standard $\ch{UO_2}$ fuel.
\newblock {\em Journal of Nuclear Materials}, 488:123--128, 2017.

\bibitem{barreiro2021aqueous}
Alexandre Barreiro-Fidalgo, Olivia Roth, Lena~Zetterstr{\"o}m Evins, and Kastriot Spahiu.
\newblock Aqueous leaching of $\ch{Cr}$ 2 o 3-doped uo 2 spent nuclear fuel under oxidizing conditions.
\newblock {\em MRS Advances}, 6:103--106, 2021.

\bibitem{smith2023oxidative}
Hannah Smith, Th{\'e}o Cordara, Cl{\'e}mence Gausse, Sarah~E Pepper, and Claire~L Corkhill.
\newblock Oxidative dissolution of $\ch{Cr}$-doped $\ch{UO_2}$ nuclear fuel.
\newblock {\em npj Materials Degradation}, 7(1):25, 2023.

\bibitem{RODRIGUEZVILLAGRA2022153880}
N.~Rodríguez-Villagra, O.~Riba, A.~Milena-Pérez, J.~Cobos, L.~Jimenez-Bonales, S.~Fernández-Carretero, E.~Coene, O.~Silva, and L.~Duro.
\newblock Dopant effect on the spent fuel matrix dissolution of new advanced fuels: $\ch{Cr}$-doped $\ch{UO_2}$ and $\ch{Cr}$ /$\ch{Al}$ -doped $\ch{UO_2}$.
\newblock {\em Journal of Nuclear Materials}, 568:153880, 2022.

\bibitem{rodriguez2025impact}
N~Rodr{\'\i}guez-Villagra, S~Fern{\'a}ndez-Carretero, A~Milena-P{\'e}rez, LJ~Bonales, L~Gutierrez, J~Cobos, and H~Gal{\'a}n.
\newblock Impact of dopants and leachants on modern $\ch{UO_2}$-based fuels alteration under final storage conditions: single and joint effects.
\newblock {\em Journal of Nuclear Materials}, page 155635, 2025.

\bibitem{lupercio2025statistical}
Adrianna~E Lupercio, Tashiema Ulrich, Andrew~T Nelson, and Brian~J Jaques.
\newblock Statistical fracture behavior of doped $\ch{UO_2}$ using a ball-on-ring equibiaxial flexure test method.
\newblock {\em Journal of Nuclear Materials}, page 155713, 2025.

\bibitem{CURTI2020152140}
Enzo Curti and Dmitrii~A. Kulik.
\newblock Oxygen potential calculations for conventional and $\ch{Cr}$-doped $\ch{UO_2}$ fuels based on solid solution thermodynamics.
\newblock {\em Journal of Nuclear Materials}, 534:152140, 2020.

\bibitem{smith3assessment}
H~Smith, T~Cordara, R~Mohun, MC~Stennett, NC~Hyatt, and CL~Corkhill.
\newblock Assessment of long-term durability of $\ch{Cr_2O_3}$ doped $\ch{UO_2}$.
\newblock In {\em 3 rd Annual Meeting Proceedings}, page~89, 2020.

\bibitem{NIELSON2024}
William~D. Neilson, Jason Rizk, Michael W.~D. Cooper, and David~A. Andersson.
\newblock Oxygen potential, uranium diffusion, and defect chemistry in $\ch{UO_2}$±x: A density functional theory study.
\newblock {\em The Journal of Physical Chemistry C}, 0(0):null, 0.

\bibitem{kardoulaki2019report}
Erofili Kardoulaki.
\newblock Report on the effect of dopants on flash sintering of enhanced $\ch{UO_2}$.
\newblock Technical report, Los Alamos National Laboratory (LANL), Los Alamos, NM (United States), 2019.

\bibitem{cordara2020hot}
Theo Cordara, Hannah Smith, Ritesh Mohun, Laura~J Gardner, Martin~C Stennett, Neil~C Hyatt, and Claire~L Corkhill.
\newblock Hot isostatic pressing (hip): a novel method to prepare $\ch{Cr}$-doped uo 2 nuclear fuel.
\newblock {\em MRS Advances}, 5:45--53, 2020.

\bibitem{VALLELY2025156105}
Sarah Vallely, Ritesh Mohun, David~W. Williams, P.~John Thomas, Mattias Puide, David~T. Goddard, William~E. Lee, and Simon~C. Middleburgh.
\newblock Optical dilatometry and sintering studies of cr2o3-doped uo2.
\newblock {\em Journal of Nuclear Materials}, 616:156105, 2025.

\bibitem{MATSUI1986212}
Tsuneo Matsui and Keiji Naito.
\newblock Electrical conductivity measurement and thermogravimetric study of chromium-doped uranium dioxide.
\newblock {\em Journal of Nuclear Materials}, 137(3):212--216, 1986.

\bibitem{GREENQUIST2020152052}
Ian Greenquist, Michael Tonks, Michael Cooper, David Andersson, and Yongfeng Zhang.
\newblock Grand potential sintering simulations of doped $\ch{UO_2}$ accident-tolerant fuel concepts.
\newblock {\em Journal of Nuclear Materials}, 532:152052, 2020.

\bibitem{greenquist2023phase}
Ian Greenquist, Amani Cheniour, Tash Ulrich, Andrew Kercher, and Ashley Shields.
\newblock Phase field sintering simulations of tagged $\ch{UO_2}$.
\newblock Technical report, Oak Ridge National Laboratory (ORNL), Oak Ridge, TN (United States), 2023.

\bibitem{yang2024novel}
Zhenliang Yang, Bingqing Li, Jingkun Xu, Yi~Zhong, Liang Xie, Mingfu Chu, Yun Wang, Rui Gao, Libing Yu, Mingshan Wang, et~al.
\newblock A novel class of atf fuels with large grain size, enhanced thermophysical properties and oxidation resistance.
\newblock {\em Ceramics International}, 50(11):18986--18992, 2024.

\bibitem{COOPER2018251}
M.W.D. Cooper, S.T. Murphy, and D.A. Andersson.
\newblock The defect chemistry of $\ch{UO_2}$±x from atomistic simulations.
\newblock {\em Journal of Nuclear Materials}, 504:251--260, 2018.

\bibitem{RONDINELLA201024}
Vincenzo~V. Rondinella and Thierry Wiss.
\newblock The high burn-up structure in nuclear fuel.
\newblock {\em Materials Today}, 13(12):24--32, 2010.

\bibitem{NOIROT2008318}
J.~Noirot, L.~Desgranges, and J.~Lamontagne.
\newblock Detailed characterisations of high burn-up structures in oxide fuels.
\newblock {\em Journal of Nuclear Materials}, 372(2):318--339, 2008.

\bibitem{zhao2024transient}
Dong Zhao, Heng Ban, Kun Yang, Andre Broussard, Mingxin Li, Edward~J Lahoda, and Jie Lian.
\newblock Transient behavior of oxide fuels with controlled microstructure and $\ch{Cr_2O_3}$ additive.
\newblock {\em npj Materials Degradation}, 8(1):74, 2024.

\bibitem{noirot2022restructuring}
Jean Noirot, R{\'e}becca Dowek, Isabelle Zacharie-Aubrun, Thierry Blay, Martiane Cabi{\'e}, and Myriam Dumont.
\newblock Restructuring in high burn-up pressurized water reactor $\ch{UO_2}$ fuel central parts: Experimental 3d characterization by focused ion beam—scanning electron microscopy.
\newblock {\em Journal of Applied Physics}, 132(19), 2022.

\bibitem{zhaoeffects}
Dong Zhao, Saurabh~Kumar Sharma, Andre Broussard, Kevin Yan, Tianyi Chen, and Jie Lian.
\newblock Effects of porosity and $\ch{Cr_2O_3}$ doping on transient behavior of high-burnup $\ch{UO_2}$ under simulated ria/loca.
\newblock {\em Journal of the American Ceramic Society}, page e20699, 2025.

\bibitem{milena2025oxidation}
Abel Milena-P{\'e}rez, Jone~M Elorrieta, Lorenza Emblico, Laura~J Bonales, Daniel Serrano-Purroy, Nieves Rodr{\'\i}guez-Villagra, and Hitos Galan.
\newblock Oxidation resistance of high-burnup $\ch{Cr}$-doped $\ch{UO_2}$ accident tolerant fuel and comparison with irradiated $\ch{UO_2}$.
\newblock {\em Journal of Nuclear Materials}, page 155930, 2025.

\bibitem{milena2025raman}
A~Milena-P{\'e}rez, LJ~Bonales, N~Rodr{\'\i}guez-Villagra, J~Cobos, and H~Gal{\'a}n.
\newblock Raman spectroscopy study of the influence of additives ($\ch{Cr}$-, $\ch{Cr}$ /$\ch{Al}$ -, and $\ch{Gd}$) on $\ch{UO_2}$ dissolution behavior.
\newblock {\em MRS Advances}, pages 1--7, 2025.

\bibitem{owen2022role}
Megan Owen.
\newblock {\em The role of alloying elements on grain boundary complexions in nuclear materials}.
\newblock Bangor University (United Kingdom), 2022.

\bibitem{BAWANE2021113790}
Kaustubh Bawane, Panayotis Manganaris, Yachun Wang, Jagadeesh Sure, Arthur Ronne, Phillip Halstenberg, Sheng Dai, Simerjeet~K. Gill, Kotaro Sasaki, Yu~chen Karen Chen-Wiegart, Ruchi Gakhar, Shannon Mahurin, Simon~M. Pimblott, James~F. Wishart, and Lingfeng He.
\newblock Determining oxidation states of transition metals in molten salt corrosion using electron energy loss spectroscopy.
\newblock {\em Scripta Materialia}, 197:113790, 2021.

\bibitem{Wolf2025}
Matthew~J. Wolf, Adrian~L. Usler, and Roger~A. De~Souza.
\newblock Grain-boundary corrosion in $\text{UO}_2$+$\delta$ from a defect chemical perspective: A case study of the $\sigma$5(310)[001] grain boundary.
\newblock {\em ACS Applied Materials \& Interfaces}, 17(5):7906--7915, 2025.
\newblock PMID: 39853164.

\bibitem{KIM2024116706}
Sangtae Kim, Sergey Khodorov, Leonid Chernyak, Thomas Defferriere, Harry Tuller, and Igor Lubomirsky.
\newblock Quantitative determination of charge trapped at grain boundaries in ionic conductors by impedance spectroscopy.
\newblock {\em Solid State Ionics}, 417:116706, 2024.

\bibitem{toyama2023real}
Satoko Toyama, Takehito Seki, Yuya Kanitani, Yoshihiro Kudo, Shigetaka Tomiya, Yuichi Ikuhara, and Naoya Shibata.
\newblock Real-space observation of a two-dimensional electron gas at semiconductor heterointerfaces.
\newblock {\em Nature Nanotechnology}, 18(5):521--528, 2023.

\bibitem{toyama2024direct}
Satoko Toyama, Takehito Seki, Bin Feng, Yuichi Ikuhara, and Naoya Shibata.
\newblock Direct observation of space-charge-induced electric fields at oxide grain boundaries.
\newblock {\em Nature Communications}, 15(1):8704, 2024.

\bibitem{Gilgenbach2024}
Colin Gilgenbach, Thomas Defferriere, Harry~L Tuller, and James~M LeBeau.
\newblock Direct quantification of grain boundary space charge layers using multislice electron ptychography.
\newblock {\em Microscopy and Microanalysis}, 30(Supplement 1):ozae044.916, 07 2024.

\bibitem{galvin2022development}
Conor Oscar~T Galvin, Aritra Chakraborty, Anders David~Ragnar Andersson, Laurent Capolungo, and Michael William~Donald Cooper.
\newblock Development of a creep model informed by lower-length scale simulations to simulate creep in doped $\ch{UO_2}$.
\newblock Technical report, Los Alamos National Laboratory (LANL), Los Alamos, NM (United States), 2022.

\end{thebibliography}

\end{document}